\documentclass[journal,twoside]{IEEEtran}
\IEEEoverridecommandlockouts                              

\usepackage{amsmath,amsfonts,amssymb}
\usepackage{cite}
\usepackage{color}
\usepackage{graphicx}
\usepackage{array}
\usepackage{textcomp}
\usepackage{siunitx}

\newcommand{\qed}{\hspace*{\fill}$\Box$\par}
\newtheorem{theorem}{Theorem}

\newtheorem{definition}{Definition}

\newtheorem{axiom}{Axiom}
\newtheorem{allocation}{Allocation}

\title{\LARGE \bf
Analysis of Solar Energy Aggregation under Various Billing Mechanisms
}

\author{Pratyush Chakraborty, Enrique Baeyens,  Pramod P. Khargonekar, Kameshwar Poolla, and Pravin Varaiya
\thanks{This research is supported by the National Science Foundation under grants EAGER-1549945, CPS-1646612, CNS-1723856 and by the National Research Foundation of Singapore under a grant to the Berkeley Alliance for Research in Singapore}
\thanks{P. Chakraborty  is with the Department of Mechanical Engineering,
        University of California, Berkeley, CA, USA
        {\tt\footnotesize pchakraborty@berkeley.edu}}%
\thanks{E. Baeyens is with Instituto de las Tecnolog\'{\i}as Avanzadas de la Producci\'on,
Universidad de Valladolid, Valladolid, Spain
        {\tt\footnotesize enrbae@eis.uva.es}}%
\thanks{K. Poolla and P. Varaiya is with the Department of Electrical Engineering and Computer Science,
        University of California, Berkeley, CA, USA
        {\tt\footnotesize poolla@berkeley.edu, varaiya@berkeley.edu}}
\thanks{P. P. Khargonekar is with the Department of Electrical Engineering and Computer Science,
        University of California, Irvine, CA, USA
        {\tt\footnotesize pramod.khargonekar@uci.edu }}%
}

\begin{document}

\maketitle
\thispagestyle{empty}
\pagestyle{empty}


\begin{abstract}
Ongoing reductions in the cost of solar photovoltaic (PV) systems are driving
their increased installations by residential households.  Various incentive programs
such as feed-in tariff, net metering, net purchase and sale that allow the
prosumers to sell their generated electricity to the grid are also powering
this trend. In this paper, we investigate sharing of PV systems among a
community of households, who can also benefit further by pooling their
production.  Using cooperative game theory, we find conditions under which such
sharing decreases their net total cost. We also develop allocation rules such
that the joint net electricity consumption cost is allocated to the
participants. These cost allocations are based on the cost causation principle.
The allocations also satisfy the standalone cost principle and promote PV solar
aggregation.  We also perform a comparative analytical study on the benefit of
sharing under the mechanisms favorable for sharing, namely net metering,
and net purchase and sale.  The results are illustrated in a case
study using real consumption data from a residential community in Austin,
Texas.

\end{abstract}

\begin{IEEEkeywords}
Community solar, Solar PV aggregation, Net metering, Net purchase and sale,
Cost allocation based on cost causation, Cooperative games.
\end{IEEEkeywords}

\section{Introduction}


Greater adoption of residential scale solar photovoltaic renewable electric
energy is a compelling engineering and sustainability objective. Powered by (a) ongoing
price reductions~\cite{barbose2016}, (b) various types of
subsidies~\cite{hughes2015}, and (c) desire to decarbonize the energy
system~\cite{kalogirou2004}, there has been dramatic increase in rooftop solar
PV installations. As residential users install rooftop solar panels and
generate portions of their electricity needs, the need for fossil fuel based
electric power plants decreases.
The total PV installations globally have reached 300 GW by 2016~\cite{ren2017}
of which about 28\% are decentralized grid connected.  We are focused on developing techniques and tools that can further increase the
cost-effectiveness of rooftop solar PV installations.

Three main billing programs around the world
enable homeowners to sell their PV electricity to the grid:
\emph{feed-in-tariff}, \emph{net metering}, and
\emph{net purchase and sale}~\cite{yamamoto2012}.
Some utilities consider these programs as a threat to their business
models~\cite{warrick2015}. On the other hand, socio-economic-environmental
policies surrounding climate change have led various governments to
encourage such programs.

In this paper, we investigate how sharing the electricity generated by rooftop
PV in a cooperative manner can further facilitate their adoption by decreasing
overall energy costs. We assume that the rooftop solar panels are electrically
connected with each other and the necessary hardware for electricity sharing
has been installed.
Sharing economy has been a huge success in housing and transportation sectors
in recent times~\cite{heinrichs2013}. It has been propelled by the desire to
leverage under-utilized infrastructures in existing houses and cars.
Companies like Uber, Lyft, AirBnB, VRBO made large impacts in transportation
and housing sectors~\cite{zervas2014}. In the electricity sector, there is some
research on modeling resource sharing. Cooperation and aggregation of
renewable energy sources bidding in the two settlement market to maximize
expected and realized profit has been analyzed using cooperative game
theory~\cite{baeyens2013,Chakraborty2016}. Cooperative game theoretic
analysis of multiple demand response aggregators in a virtual power plant and
their cost allocation has been tackled in~\cite{nguyen2017}. Sharing of storage
firms under a local spot market has been analyzed using non-cooperative game
theory~\cite{kalathil2017}.

Community solar projects, where a community scale solar energy plant is developed and shared by many households, are getting popular \cite{energysage}. Various countries like Germany, Denmark, Australia, United Kingdom, United States are already investing significantly in this type of projects \cite{Ison2015}. Benefit of shared solar PV under net metering has been studied in \cite{funkhouser2015}. To the best of our knowledge, sharing of PV systems in a cooperative manner among different houses
under various billing schemes has not been investigated.
Can cooperation among rooftop PV systems reduce costs to prosumers under all
billing mechanisms? If the answer is affirmative, to encourage and preserve
cooperative sharing, it will be crucial to have a just and reasonable
allocation of the resulting cost reduction or benefit increase to the
participating individuals. These are precisely the questions we analyze in
this paper.

Our results show that there is no advantage for cooperation in the case of
feed-in tariff. Using cooperative game theory, we derive a necessary
and sufficient condition on pricing under which cooperation is advantageous
for the participating prosumers in net metering, and net purchase and sale
mechanisms. We develop rules for allocating joint cost based on the
cost causation principle~\cite{Chakraborty2017}.
We prove that these cost allocation rules also follow standalone
cost principle --\emph{i.e.}, they are in the core of the
cooperative games of net electricity consumption costs.
Next, we conduct a
comparative study about the cooperation benefits under net metering,
and net purchase and sale mechanisms.
These results provide theoretical basis for sharing of rooftop PV generation
among prosumers. We also develop a case study based on real consumption and
generation data of a residential community in Austin, Texas, to illustrate
our results.

The remainder of the paper is organized as follows. In Section~\ref{sec:PF},
we formulate the problem, introduce notation and describe the system model.
In Section~\ref{sec:PR}, we review basic results on cooperative game theory
and cost allocation based on cost causation that are used in the derivation of the results.
In Section~\ref{sec:R}, we report our main results on aggregation of PV
solar energy. A case study to illustrate the theoretical results is
presented in Section~\ref{sec:NE}. Finally, we conclude our work in
Section~\ref{sec:Con}.

\section{Problem Formulation}
\label{sec:PF}

Consider a residential community of $N$ households with PV systems, that
is represented by the set $\mathcal N=\{1,2,\ldots,N \}$.
The retail price of electricity consumption for the set of households
is $\lambda$. They can also sell their electricity generation at price
$\mu$. Both $\lambda$ and $\mu$ are decided by the utility operating under
the system operators based on different factors, policies and countries regulations~\cite{yamamoto2012}. We consider three programs:
feed-in tariff, net metering, and net purchase and sale that
allow households to sell their generated electricity to the
grid. These three mechanisms are explained in detail in~\cite{yamamoto2012},
indicating their countries of usage and also analyzing their impact on social
welfare. In the sequel, and in order to improve clarity of the presentation,
we shall consider that the prices $\lambda$ and $\mu$ do not change
during each billing period. However, our results could be easily extended
to an scenario of either time-of-use (TOU) tariffs or dynamic prices at the
expense of a burdensome notation.
We also assume here that the households obtain more utility than
their cost for solar generation and thus rationally
install solar panels.

In \emph{feed-in tariff} program, the households can sell all of their PV
generation at price $\mu$ and they must purchase all of their consumed
electricity from the grid at price $\lambda$. In contrast, under the programs
net metering, and net purchase and sale, electric utility of
the grid purchases only the net amount of the PV generation of the households
that exceeds their consumption. But the two programs compute their net
electricity consumption and generation in different ways.
Under \emph{net metering} program, when the PV generation of a household exceeds
its consumption, the electric meter runs backwards. At the end of a billing
period, if the amount of electricity generation is more than consumption, the
household is paid for the net PV generation at price $\mu$. If the amount of
electricity generation is less than consumption, the household has to pay the
net amount consumed at price $\lambda$.
Under \emph{net purchase and sale} program, the generation and consumption
is compared at each time and the PV generated electricity is fed into the
system actually when generation exceeds consumption and purchased by the
utility at price $\mu$. Otherwise, the consumed electricity is purchased by
the household at price $\lambda$. So the amount of electricity is compared
moment by moment in this program instead of at the end of a billing period
as in the net metering program.

Let us consider a billing period $[t_0,t_f]$ of duration $T=t_f-t_0$.
For the $i$-th household at time $t \in [t_0,t_f]$, let the electricity
consumption be $q_i(t)$, and the generation by rooftop solar panels
be $g_i(t)$.

The net electricity consumption cost of the household for the entire billing
period is:
\begin{equation}\label{eq:cost}
C_i^b = \lambda Q_i^b  - \mu G_i^b,
\end{equation}
where $Q_i^b$ and $G_i^b$ denote the $i$-th household's energy
consumption and generation during the billing period for the billing
mechanism $b$, respectively.
These quantities are computed in a different way depending on the
billing mechanism.

For \emph{feed-in tariff} mechanism ($b=\text{FiT}$):
\begin{align}
Q_i^{\text{FiT}} = \int_{t_0}^{t_f} q_i(t) dt, \quad
G_i^{\text{FiT}} = \int_{t_0}^{t_f} g_i(t) dt. \label{eq:QGifit}
\end{align}

For \emph{net metering} mechanism ($b=\text{NM}$):
\begin{align}
Q_i^{\text{NM}} &=
\left(\int_{t_0}^{t_f} q_i(t) dt - \int_{t_0}^{t_f} g_i(t) dt\right)_+,
\label{eq:Qinm} \\
G_i^{\text{NM}}
&= \left(\int_{t_0}^{t_f} g_i(t) dt - \int_{t_0}^{t_f} q_i(t) dt\right)_+.
\label{eq:Ginm}
\end{align}

For \emph{net purchase and sale} mechanism ($b=\text{NPS}$):
\begin{align}
Q_i^{\text{NPS}}
&= \int_{t_0}^{t_f} (q_i(t) - g_i(t))_+ dt, \label{eq:Ginps} \\
G_i^{\text{NPS}}
&= \int_{t_0}^{t_f} (g_i(t) - q_i(t))_+ dt. \label{eq:Qinps}
\end{align}

Note that the net consumption for each household $i\in\mathcal N$
is obtained as
\begin{align}
D_i=Q_i^b-G_i^b,
\end{align}
and has the same value for the three billing mechanisms,
but the cost $C_i^b$ is different and depends on the billing mechanism
$b \in \{\text{FiT},\text{NM},\text{NPS}\}$.

Let $\mathcal S \subseteq \mathcal N$ denote a coalition of households that
decide to cooperate to share their electricity generation by rooftop solar
panels and save electricity costs.  We assume that the rooftop solar panels are
electrically connected with each other and the houses also have necessary
hardwares for electricity sharing.  The total energy consumption and generation
of the coalition for the interval $[t_0,t_f]$ and the billing mechanism $b$ are
denoted as $Q_{\mathcal S}^b$ and $G_{\mathcal S}^b$, respectively.
The consumption is charged at price $\lambda$ and the generation is paid
at price $\mu$, which are assumed to be constant during the billing period.
Consequently, the cost of the coalition for the time interval $[t_0,t_f]$ is
\begin{align}
C_{\mathcal S}^b &=\lambda Q_{\mathcal S}^b - \mu G_{\mathcal S}^b,
\end{align}
where the expressions of $Q_{\mathcal S}^b$ and $G_{\mathcal S}^b$ depend on
the billing mechanism $b \in \{\text{FiT},\text{NM},\text{NPS}\}$.

Cooperation is advantageous for the set $\mathcal N$ of households under
the billing mechanism $b$ if the joint electricity cost
$C_{\mathcal S}^b$ is a subadditive function, \emph{i.e.} if for any
any pair of disjoint coalitions
$\{(\mathcal S, \mathcal T):\mathcal S, \mathcal T \subseteq \mathcal N\}$,
$C_{\mathcal S \cup \mathcal T}^b \leq C_{\mathcal S}^b + C_{\mathcal T}^b$.
We are interested in studying under which billing mechanisms and what conditions,
cooperation is advantageous. Moreover, if cooperation produce some benefit,
we want to develop mechanisms to allocate it among the households in a way
that is satisfactory for all of them. These questions can be formally
answered using cooperative game theory.

\section{Background on Cooperative Game Theory for Cost Sharing and Allocation}
\label{sec:PR}

Game theory deals with rational behavior of economic agents in a mutually
interactive setting \cite{Neumann1944}.
Cooperative games (or coalitional games) have been used extensively in diverse
disciplines such as social science, economics, philosophy,
psychology and communication networks \cite{Myerson2013,Saad2009}.
Here, we focus on cooperative games for cost sharing \cite{Jain2007}. We
explain the theory with the setup and notation of our solar PV aggregation
problem.

Let $\mathcal{N}:=\{1,2,\ldots,N\}$ denote a finite collection of players. In a cooperative game for cost sharing, the players want to minimize their joint cost and share the resulting cost cooperatively.
\begin{definition}[Coalition]
A \emph{coalition} is any subset $\mathcal S \subseteq \mathcal N$.
The number of players in a coalition $\mathcal S$ is denoted by its cardinality, $|\mathcal S|$.
The \emph{set of all possible coalitions} is defined as the power set $2^{\mathcal N}$ of $\mathcal N$.
The \emph{grand coalition} is the set of all players, $\mathcal N$.
\end{definition}

\begin{definition}[Game and Cost]
A cooperative game is defined by a pair $(\mathcal{N},C)$ where
$C:2^{\mathcal{N}}\rightarrow \mathbb R$
is the \emph{cost function} that assigns a real value to each coalition
$\mathcal{S} \subseteq \mathcal{N}$.
Hence, the \emph{cost of the coalition} $\mathcal S$ is given by
$C_{\mathcal S}=C(\mathcal S)$.
\end{definition}

\begin{definition}[Subadditive Game]
A cooperative game $(\mathcal N, C)$ is \emph{subadditive} if, for any pair of disjoint coalitions
$\mathcal S, \mathcal T \subset \mathcal N$ with
$\mathcal S \cap \mathcal T = \emptyset$, we have $C_{\mathcal S} + C_{\mathcal T} \ \geq \ C_{\mathcal S \cup  \mathcal T}.$
\end{definition}

Here we consider the value of the coalition $C_{\mathcal S} $ is
\emph{transferable} among players.  The central question for a subadditive cost
sharing game with transferrable value is how  to \emph{fairly} distribute the
coalition value among the coalition members.

\begin{definition}[Cost Allocation]
A \emph{cost allocation} for the coalition $\mathcal{S} \subseteq \mathcal N$
is a vector $x\in\mathbb{R}^{N}$
whose entry $x_i$ represents the allocation to member $i \in \mathcal S$
($x_i = 0, \ \ i \notin \mathcal S$).
\end{definition}
For any coalition $\mathcal S\subseteq{\mathcal N}$, let $x_{\mathcal S}$
denote the sum of cost allocations for every coalition member, \emph{i.e.}
$x_{\mathcal S}=\sum_{i\in\mathcal S}x_i$.

\begin{definition}[Imputation]
A cost allocation $x$ for the grand coalition $\mathcal N$ is said to be an
{\em imputation} if it is simultaneously budget balanced
--\emph{i.e.} $C_{\mathcal N}=\sum_{i=1}^{N} x_i$, and individually rational
--\emph{i.e.} $C_{i}\geq x_i, \forall i \in \mathcal N$.
Let $\mathcal I$ denote the set of all imputations.
\end{definition}

The fundamental solution concept for cooperative games is the \emph{core}
\cite{Neumann1944}.

\begin{definition}[The Core]  \label{core_def}
The \emph{core} $\mathcal{C}$ for the cooperative game $(\mathcal{N},C)$
with \emph{transferable cost} is defined as the set of cost allocations
such that no coalition can have cost which is lower than the
sum of the members current costs under the given allocation.
\begin{align}
\label{core_eq}
\mathcal C :=
\left\{
x \in \mathcal I:
C_\mathcal S\geq x_{\mathcal S},
\forall \mathcal S \in 2^{\mathcal N}
\right\}.
\end{align}
\end{definition}

A classical result in cooperative game theory, known as Bondareva-Shapley theorem, gives a necessary and sufficient condition for a game to have nonempty core. To state this theorem, we need the following definition.

\begin{definition}[Balanced Game and Balanced Map]
A cooperative game $(\mathcal N, C)$ for cost sharing is \emph{balanced} if for any balanced map
$\alpha$,
$\sum_{\mathcal S\in 2^{\mathcal N}}
\alpha(\mathcal S) C_{\mathcal S} \geq C_{\mathcal N}$ where the map
$\alpha: 2^{\mathcal N} \rightarrow [0,1]$ is said to be \emph{balanced}
if for all $i\in\mathcal N$, we have $\sum_{\mathcal S\in 2^{\mathcal N}}
\alpha(\mathcal S) \mathbf{1}_{\mathcal S}(i) = 1$, where
$\mathbf{1}_{\mathcal S}$ is the indicator function of the set $\mathcal S$,
\emph{i.e.} $\mathbf{1}_{\mathcal S}(i)=1$ if $i\in\mathcal S$ and
$\mathbf{1}_{\mathcal S}(i)=0$ if $i\not\in\mathcal S$.
\end{definition}
Next we state the Bondareva-Shapley theorem.

\begin{theorem}[Bondareva-Shapley Theorem \cite{Saad2009}]
A coalitional game has a nonempty core if and only if it is balanced.
\end{theorem}

There are a number of solutions that exist for a cooperative game.
The most prominent are the \emph{Shapley value}, the
\emph{nucleolus} \cite{Myerson2013} and the minimum worst-case excess
allocation~\cite{baeyens2013}.
If a game is balanced, the nucleolus and the minimum worst-case excess
are always in the core. For a concave cost sharing game, the Shapley
value is also in the core. These allocations are computationally
costly for large number of players. In order to circumvent
this problem and also to analyse properties of an allocation, we proposed in \cite{Chakraborty2017} an axiomatic framework
to characterize \emph{just and reasonable} cost allocation rules.
These axioms are: \emph{equity}, \emph{monotonicity},
\emph{individual rationality}, \emph{budget balance}, and
\emph{standalone cost principle}. These axioms are established using a
variable that characterizes the cause of cost of an agent.
For the problem of allocation of the aggregated cost of
a group of households, the net individual consumption $D_i=Q_i^b-G_i^b$
for $i\in\mathcal N$ is a natural choice for the variable that characterizes
the cause of cost. Then, we establish the following axioms.

\begin{axiom}[Equity] \label{ax:Eq}
If two agents $i$ and $j$ have same net consumptions,
the allocated costs must be the same \emph{i.e.}, if
$D_i = D_j$ then $x_i=x_j$.
\end{axiom}

\begin{axiom}[Monotonicity] \label{ax:Mo}
If two agents $i$ and $j$ have net consumptions of the same sign,
and agent $i$ has a higher net consumption than agent $j$,
then the absolute value of the allocated cost to $i$ must be higher than
the absolute value of the allocated cost to $j$ \emph{i.e.},
if $D_iD_j\geq 0$ and
$|D_i| \geq |D_j|$ then
$|x_i| \geq |x_j|$.
\end{axiom}

\begin{axiom}[Individual Rationality] \label{ax:IR}
The allocated cost must be less than the cost
if the agent would not have joined the aggregation
\emph{i.e.}, $x_i\leq C_i$.
\end{axiom}

\begin{axiom}[Budget Balance] \label{ax:BB}
The cost allocation rule would be such
that the sum of allocated costs must be equal to the net electricity consumption cost
\emph{i.e.}, $\sum_{i\in N} x_i= C_\mathcal{N}.$
\end{axiom}

\begin{axiom}[Standalone Cost Principle] \label{ax:ST}
For every coalition $\mathcal S \subset \mathcal{N}$,
\begin{align}
x_{\mathcal S}\leq C_{\mathcal S}.
\end{align}
\end{axiom}
From Axiom \ref{ax:ST} and Definition \ref{core_def}, it is easy to see that
an allocation that satisfies
the standalone cost principle belongs to the core of the cooperative game of cost sharing.

A cost causation based allocation was inspired by the tariffs
proposed by Kirby in \cite{Kirby2006} and should follow the axioms of
equity, monotonicity, individual rationality and budget balance, but
not necessarily the standalone cost principle.
However, not every allocation rule satisfying
these four axioms follows the cost causation principle, because
they do not explicitly take into account whether
agents are causing or mitigating costs.

Using again the net individual consumption $D_i$ for $i\in\mathcal N$
as the variable signaling cause or mitigation of cost,
we state the following cost causation axioms.

\begin{definition}[Cost causation and mitigation] \label{def:CCM}
Let $D_{\mathcal N}$ be the net consumption
of a group $\mathcal N$ of agents. It is said that agent $i$ is
causing cost if $D_i > 0$, and is mitigating cost if $D_i < 0$.
\end{definition}

Based on this definition, we introduce two new cost causation based axioms:
\emph{penalty for cost causation} and \emph{reward for cost mitigation}.

\begin{axiom}[Penalty for causing cost] \label{ax:cc1}
Those individuals causing cost should pay for it,
\emph{i.e.} $x_i > 0$ for any $i\in\mathcal N$
such that $D_i > 0$.
\end{axiom}

\begin{axiom}[Reward for cost mitigation] \label{ax:cc2}
Those individuals mitigating cost should be rewarded,
\emph{i.e.} $x_i < 0$ for any $i \in \mathcal N$
such that $D_i < 0$.
\end{axiom}

Using the previously introduced axioms, a \emph{cost causation based}
allocation rule is formally defined as follows.

\begin{definition}[Cost causation based allocation rule] \label{def:CCBA}
A cost allocation rule is said to be a \emph{cost causation based allocation}
rule if it satisfies Axioms \ref{ax:Eq}--\ref{ax:BB} and
\ref{ax:cc1}--\ref{ax:cc2}.
\end{definition}

It is interesting to remark that some well-known allocation rules such
as the \emph{proportional rule}, or the \emph{Shapley value}, do not
satisfy the cost causation axioms \cite{Chakraborty2017}.

\section{Main results on solar energy aggregation and cost allocation under
different billing mechanisms}
\label{sec:R}

In this section, we present the main results about cooperation of
households for different billing programs. We study the conditions
for each billing program to provide successful cooperation.

\subsection{Feed-in tariff}

For the coalition $\mathcal S\subseteq\mathcal N$,
under the feed-in tariff program, the net consumption
during the billing period $[t_0,t_f]$ is
\begin{align}
D_{\mathcal S}=Q_{\mathcal S}^{\text{FiT}} - G_{\mathcal S}^{\text{FiT}},
\end{align}
where
\begin{align}
Q_{\mathcal S}^{\text{FiT}} = \sum_{i \in \mathcal S} Q_i^{\text{FiT}}, \quad
G_{\mathcal S}^{\text{FiT}} = \sum_{i \in \mathcal S} G_i^{\text{FiT}},
\end{align}
and $Q_i^{\text{FiT}}$, $G_i^{\text{FiT}}$ are given by equations
(\ref{eq:QGifit}).

Let us consider two disjoint coalitions $\mathcal S$ and $\mathcal T$.
For this billing program, the cost of the coalition $\mathcal S$ is
\begin{equation}
C_{\mathcal S}^{\text{FiT}}=
\lambda Q_{\mathcal S}^{\text{FiT}} -\mu G_{\mathcal S}^{\text{FiT}}.
\end{equation}
It is easy to see that
$C_{\mathcal S}^{\text{FiT}}+C_{\mathcal T}^{\text{FiT}}=
C_{\mathcal S \cup \mathcal T}^{\text{FiT}}$.
So cooperation is neutral and there is no advantage or harm in cooperating and
sharing the PV generation. However, since there is likely to be some cost for enabling cooperation, this result suggests that FiT will not encourage cooperation.

\subsection{Net Metering}

The aggregated consumption and generation for any coalition
$\mathcal S \subseteq \mathcal N$ during the billing period
$[t_0,t_f]$ under the net metering program are given by
\begin{align}
Q_{\mathcal S}^{\text{NM}} &=
\left(\sum_{i \in \mathcal S}\int_{t_0}^{t_f} q_i(t) dt  -
\sum_{i \in \mathcal S}\int_{t_0}^{t_f} g_i(t) dt  \right)_+,\\
G_{\mathcal S}^{\text{NM}} &=
\left(\sum_{i \in \mathcal S}\int_{t_0}^{t_f} g_i(t) dt  -
\sum_{i \in \mathcal S} \int_{t_0}^{t_f} q_i(t) dt \right)_+.
\end{align}
The net consumption of the coalition is
\begin{align}
D_{\mathcal S}=Q_{\mathcal S}^{\text{NM}} - G_{\mathcal S}^{\text{NM}},
\end{align}
and the joint cost of the net consumption for the coalition is
\begin{equation}
C_{\mathcal S}^{\text{NM}}=
\lambda Q_{\mathcal S}^{\text{NM}} - \mu G_{\mathcal S}^{\text{NM}}.
\end{equation}

Let $(\mathcal N, C^{\text{NM}})$ denote the cooperative game for the set of
of households under the net metering billing program. In the following theorem, we obtain a necessary and sufficient
condition for the game $(\mathcal N, C^{\text{NM}})$ to be
subadditive and balanced.

\begin{theorem}\label{th:nmrsc}
The cooperative game $(\mathcal N,C^{\text{NM}})$ is subadditive
and balanced if and only if $\lambda \geq \mu$.
\end{theorem}
\paragraph*{Proof} See Appendix.
\qed

\emph{Remark 1:}
The condition $\lambda \geq \mu$ developed in Theorem~\ref{th:nmrsc} requires that the selling price of the
generated electricity is always less that the retail price of the
electricity. This price condition is indeed realistic because the price of
consuming electricity should include generation, transmission and distribution
costs. But transmission and distribution costs should not be included in
the price that the utility pays for the energy generated by a prosumer
having solar PV systems~\cite{EEI2016}.

\emph {Remark 2:} Subadditivity of the game means that the households may obtain benefit
by aggregation, because the aggregated cost is always less than or equal to
the sum of the individual costs. Balancedness means that the core of the
game is nonempty and there exists a cost allocation of the aggregated cost
that is satisfactory for every member of the grand coalition. However,
the existence of such an allocation does not imply that it can be easily
computed. Under the pricing condition of Theorem~\ref{th:nmrsc}, we propose the following easy to compute cost allocation for
grand coalition participants.

\begin{allocation}[Net Metering] \label{alloc:nmet}
The allocated cost during the billing interval $[t_0,t_f]$
for the $i$-th household of a residential community
$\mathcal N$ that share energy under the net metering program is given by
\begin{align} \label{eq:alloc1}
x_i^{\text{NM}} &=
\left\{
\begin{array}{l}
\lambda D_i, \quad \mathrm{if} \ D_{\mathcal N} \geq 0,\\
\mu D_i, \quad \mathrm{if} \ D_{\mathcal N} < 0,
\end{array}
\right.
\end{align}
for every $i\in\mathcal N$.
\end{allocation}

In Theorem~\ref{th:nmrcc}, we prove that Allocation~\ref{alloc:nmet}
is a cost causation based cost allocation --\emph{i.e.}, it satisfies
equity, monotonicity, individual rationality, budget balance,
penalty for causing cost and reward for cost mitigation axioms.

\begin{theorem}\label{th:nmrcc}
The allocation defined by (\ref{eq:alloc1}) is a cost causation
based allocation.
\end{theorem}

\paragraph*{Proof} See Appendix.
\qed

In addition, Allocation~\ref{alloc:nmet} satisfies the standalone
cost principle. This means that it is, in fact, a cost allocation
in the core of the cooperative game, and no rational household has an incentive to defect from
the grand coalition, because in that case its electricity bill
would increase. The result is stated as the following theorem.

\begin{theorem}\label{th:nmrc}
The allocation defined by (\ref{eq:alloc1}) satisfies standalone
cost principle.
\end{theorem}
\paragraph*{Proof} See Appendix.
\qed

\subsection{Net Purchase and Sale}

The cost of the aggregated net consumption of a coalition
$\mathcal S \subseteq \mathcal N$ during the billing period $[t_0, t_f]$
under the net purchase and sale program is
\begin{align}
C_{\mathcal S}^{\text{NPS}}
= \int_{t_0}^{t_f} C_{\mathcal S}^{\text{NPS}}(t) dt.
\end{align}
The instantaneous cost function $C_{\mathcal S}^{\text{NPS}}(t)$
is defined as follows:
\begin{align}
C_{\mathcal S}^{\text{NPS}}(t)
&= \lambda Q_{\mathcal S}^{\text{NPS}}(t)-\mu G_{\mathcal S}^{\text{NPS}}(t),
\end{align}
where
\begin{align}
Q_{\mathcal S}^{\text{NPS}}(t) &=
(\sum_{i\in\mathcal S} q_i(t) - \sum_{i\in\mathcal S} g_i(t))_+, \\
G_{\mathcal S}^{\text{NPS}}(t) &=
(\sum_{i\in\mathcal S} g_i(t) - \sum_{i\in\mathcal S} q_i(t))_+,
\end{align}
and the instantaneous net consumption at time $t$ is
\begin{align}
D_{\mathcal S}(t)=
Q_{\mathcal S}^{\text{NPS}}(t)-G_{\mathcal S}^{\text{NPS}}(t).
\end{align}

The net purchase and sale is equivalent to the net metering where
the billing period reduces to the time instant $t \in [t_0,t_f]$. Thus,
for this billing mechanism, the study has to be accomplished instantaneously,
instead of after completing the billing period $[t_0,t_f]$.
Consequently, the definitions and axioms of Section~\ref{sec:PR}
are also valid, but they should be considered for each time instant
$t\in [t_0,t_f]$.

Let $(\mathcal N, C^{\text{NPS}})$ denote the cooperative game for the set
of households and the billing period $[t_0,t_f]$ under the
net purchase and sale program. A necessary and sufficient condition for the successful cooperation of
the group of households is given in the following theorem.

\begin{theorem}\label{th:npssc}
The cooperative game $(\mathcal N,C^{\text{NPS}})$ is subadditive and
balanced if and only if $\lambda \geq \mu$.
\end{theorem}

\paragraph*{Proof} See Appendix.

Motivated by the properties of Allocation~\ref{alloc:nmet} for the
net metering case, we propose the following cost allocation for
the net purchase and sale program.

\begin{allocation}[Net Purchase and Sale] \label{alloc:npur}
The allocated cost during the the billing time interval $[t_0,t_f]$
for the $i$-th household of a residential community
$\mathcal N$ that share energy under the net purchase and sale program
in given by
\begin{align}
\label{eq:alloc2a}
x_i^{\text{NPS}} = \int_{t_0}^{t_f} x_i^{\text{NPS}}(t) dt,
\end{align}
where
\begin{align}
\label{eq:alloc2b}
x_i^{\text{NPS}}(t) &= \left\{
\begin{array}{l}
\lambda D_i(t), \quad \mathrm{if} \ D_{\mathcal N}(t) \geq 0, \\
\mu D_i(t), \quad \mathrm{if} \ D_{\mathcal N}(t) \leq 0.
\end{array}
\right.
\end{align}
\end{allocation}

The properties of Allocation~\ref{alloc:npur} are established  in
the following theorem.

\begin{theorem}\label{th:npscc}
The allocation defined by equations (\ref{eq:alloc2a})--(\ref{eq:alloc2b})
is a cost causation based allocation that belongs to the core of the
cooperative game $(\mathcal N, C^{\text{NPS}})$.
\end{theorem}

\paragraph*{Proof} See Appendix.

\subsection{Comparative Study of Billing Mechanisms}
\label{subsec:comp}

Previously in this section, we proved that net metering, and net purchase
and sale billing mechanisms promote solar PV aggregation and thus,
renewable energy integration. Here, we are interested in quantifying
which of these mechanisms is better from the point of view of the
total cost that the prosumers pay and save as a result of sharing.

We show that the prosumers pay less under a net metering program
than under a net purchase and sale program with the same prices
$\lambda$ and $\mu$. The result is given in the
following theorem.

\begin{theorem} \label{th:difcost}
Let $\mathcal S \subseteq \mathcal N$ be any coalition of prosumers.
Assume that $\lambda$ and $\mu$ with $\lambda \geq \mu$ are the
price of purchasing and selling electricity. The cost of
the electricity consumption of the coalition is less under a net
metering program than under a net purchase and sale program, and
the difference is given by
\begin{align}
C_{\mathcal S}^{\text{NPS}} - C_{\mathcal S}^{\text{NM}} =
\left\{
\begin{array}{ll}
(\lambda-\mu) G_{\mathcal S}^{\text{NPS}}, & \text{if } D_{\mathcal S} \geq 0,\\
(\lambda-\mu) Q_{\mathcal S}^{\text{NPS}}, & \text{if } D_{\mathcal S} < 0.
\end{array}
\right.
\end{align}
\end{theorem}
\paragraph*{Proof} See Appendix.
\qed

The result is valid for any coalition and therefore it applies also to
individual prosumers. Next, we are interested in studying
if one of these programs is better in savings than the other under aggregation. We show that,
under certain conditions, the net purchase and sale program produces a
larger cost savings than the net metering program.

The aggregated cost of the grand coalition $\mathcal N$
under net metering program is:
\begin{align}
C_{\mathcal N}^{\text{NM}} =
\lambda Q_{\mathcal N}^{\text{NM}} -
\mu G_{\mathcal N}^{\text{NM}},
\end{align}
and the cost allocated to the household  $i\in\mathcal N$ using
Allocation~\ref{alloc:nmet} is:
\begin{align}
x^{\text{NM}}_i &=
\left\{
\begin{array}{l}
\lambda D_i, \quad \mathrm{if} \ D_{\mathcal N} \geq 0,\\
\mu D_i , \quad \mathrm{if} \ D_{\mathcal N} < 0.
\end{array}
\right.
\end{align}

The cost of the household $i\in\mathcal N$ when it is not sharing
solar energy with the other households is:
\begin{align}
C^{\text{NM}}_i &=
\left\{
\begin{array}{l}
\lambda D_i, \quad \mathrm{if} \ D_i \geq 0,\\
\mu D_i , \quad \mathrm{if} \ D_i < 0.
\end{array}
\right.
\end{align}

Thus, the cost saving that household $i\in\mathcal N$ obtains by
aggregating with the other households in the community $\mathcal N$ for the
billing period $[t_0,t_f]$ is
\begin{align}\label{eq:savnm}
S_i^{\text{NM}} &=
C^{\text{NM}}_i - x_i^{\text{NM}} \nonumber \\
&=
\left\{
\begin{array}{ll}
0, & \text{if } D_iD_{\mathcal N} \geq 0, \\
(\lambda-\mu)|D_i|, & \text{if } D_iD_{\mathcal N} < 0.
\end{array}
\right.
\end{align}

An analogous analysis for the net purchase and sale program provides
the instantaneous cost saving that the household $i\in\mathcal N$
obtains at time instant $t \in [t_0,t_f]$ by aggregation:
\begin{align}
S_i^{\text{NPS}}(t) &=
C^{\text{NPS}}_i(t) - x_i^{\text{NPS}}(t) \nonumber \\
&=
\left\{
\begin{array}{ll}
0, & \text{if } D_i(t)D_{\mathcal N}(t) \geq 0, \\
(\lambda-\mu)|D_i(t)|, & \text{if } D_i(t)D_{\mathcal N}(t) < 0.
\end{array}
\right.
\end{align}

For each household $i\in\mathcal N$, we define the time sets:
\begin{align}
\Omega_i^+ &= \{ t \in [t_0,t_f] : D_i(t)D_{\mathcal N}(t) \geq 0 \}, \\
\Omega_i^- &= \{ t \in [t_0,t_f] : D_i(t)D_{\mathcal N}(t) < 0 \}.
\end{align}

Then the cost saving that the household $i\in\mathcal N$ obtains by
aggregation for the billing period $[t_0,t_f]$ is:
\begin{align}
S_i^{\text{NPS}} &=
\int_{t_0}^{t_f} S_i^{\text{NPS}}(t) dt \nonumber \\
&=
\int_{\Omega_i^+} S_i^{\text{NPS}}(t)  dt +
\int_{\Omega_i^-} S_i^{\text{NPS}}(t)  dt \nonumber \\
&=
(\lambda-\mu) \int_{\Omega_i^-} |D(t)|  dt.
\end{align}

The difference in the cost savings for the household $i\in\mathcal N$
under the two billing mechanisms is given by
\begin{align}\label{eq:savcom}
& S_i^{\text{NPS}} - S_i^{\text{NM}} = \nonumber \\
& \quad
\left\{
\begin{array}{ll}
(\lambda-\mu) \int_{\Omega_i^-} |D(t)|  dt, &
\text{if } D_iD_{\mathcal N} \geq 0, \\
(\lambda-\mu) \left( \int_{\Omega_i^-} |D(t)|  dt - |D_i| \right), &
\text{if } D_iD_{\mathcal N} < 0.
\end{array}
\right.
\end{align}

Using the equations (\ref{eq:savnm}) and (\ref{eq:savcom}), we obtain that
$S_i^{\text{NPS}} \geq  S_i^{\text{NM}} = 0$ for each billing period where
$D_iD_{\mathcal N}\geq 0$. This means that, in this case,
cooperation is only advantageous under the net purchase and sale program.
However, for a billing period where the individual net consumption and the
net consumption of the grand coalition have opposite
sign --\emph{i.e.} $D_iD_{\mathcal N}\le 0$, we cannot say anything about
which program produces a greater cost saving for a household, and it depends
on the particular consumption patterns.
In a practical scenario considering household's electricity requirements,
rooftop solar capacity, solar insolation, etc., for most of the billing
periods, the households have positive individual net consumptions
--\emph{i.e.} they consume more electricity than they generate. As a result,
the grand coalition also has a positive net consumption. Thus, for most of the
billing periods and most of the households, the net purchase and sale program
produce greater cost savings than the net metering program.

\subsection{Billing Mechanisms Around The World}

Different countries use different billing mechanisms to promote
solar PV adoption~\cite{yamamoto2012}. Germany adopted feed-in-tariff, which
is not advantageous for prosumers with sharing PV systems.
Most states in the USA use net metering with $\lambda=\mu$. Japan uses net purchase and sale but in
their case $\mu \geq \lambda$. None of these billing mechanisms are
advantageous for sharing.
A high selling price $\mu \geq \lambda$ makes solar PV very attractive for
prosumers because it produces large reduction in electricity bills. However,
this policy is harmful for utilities that have to buy the generated
energy by the households at a higher price than the market price as
it includes transmission and distribution costs~\cite{EEI2016}.
It is expected that as the number of households with
rooftop solar PV increases, the selling price of the PV generated electricity
approaches the price that the utility pays for buying electricity in the
market minus the transmission and distribution costs.
Thus, in future, $\mu$ will be less than the retail price of
electricity $\lambda$, and this will promote aggregation as a means to
reduce consumption cost and integrate more renewable energy by taking
advantage of the solar PV excess.

Recently, legislators of Nevada introduced a new compensation \cite{NEV} for
excess solar energy exported to the grid at 95 percent of the retail
electricity rate. For every 80 megawatts of solar deployed, the export credit
is set to decline by 7 percent, to a floor of 75 percent of the retail rate. NV
Energy also sought to have the 95 percent calculation apply to every
kilowatt-hour of energy a prosumer exports to the grid, rather than the monthly
netting amount, a change that decreases the value prosumers get for the energy
they export. It is favorable for cooperation as $\lambda$ is greater than $\mu$. Also the
scheme reduces the value for the prosumer as calculations does not wait for
monthly netting amount, rather depends upon some energy consumption. The
mechanism is closer to net purchase and sale and as per our
analysis in this section, the sharing PV systems in this mechanism might probably give more
overall advantage to prosumers compared to net metering, though the real advantage will depend on net
consumption of a particular prosumer, as well as that of the whole cooperative
group.

\section{A case study}
\label{sec:NE}

We consider a community of \num{80} households, located in
a residential area of Austin, TX, that have PV rooftop
panels and decide to share their generation. The consumption and
generation data have been obtained from the Pecan Street project
\cite{Pecan}.
The codes of the \num{80} prosumers selected for this
study are given in Table~\ref{tab:codes}.
For each prosumer we have retrieved real data of
power consumption and solar power generation for every 15 minutes.
The period under study is the complete year of 2016, and the billing
period is one month. As we have analytically proved that there is no advantage
of sharing under feed-in tariff, in this study, we only analyze the impact on
cost and savings in sharing PV systems under net metering, and net purchase and
sale.

\subsection{Data Analysis}
We have considered
$\lambda = \textrm{\textcent}\num{11.02}/\si{\kW}$ and
$\mu = \num{0.57}\lambda$\footnote{$\lambda$ is
the average retail price of electricity in Texas during 2016, obtained
from the Energy Information Administration (EIA)
Data Browser \cite{eiaDB},
and $\mu$ corresponds exclusively to the generation part
of the retail price of electricity in the US according to the
\emph{Annual Energy Outlook 2017} \cite{EIA2017}.}.


\bgroup
\renewcommand{\tabcolsep}{1ex}
\begin{table}
\caption{Codes of the 80 prosumers used in the study}
\label{tab:codes}
\begin{tabular}{cccccccccc}
\hline
\multicolumn{10}{c}{\bf Prosumers' Codes} \\
\hline
26& 77& 93& 171& 370& 379& 545& 585& 624& 744\\
781& 890& 1283& 1415& 1697& 1792& 1800& 2072& 2094& 2129\\
2199& 2233& 2557& 2818& 2925& 2945& 2980& 3044& 3310& 3367\\
3456& 3482& 3538& 3649& 4154& 4352& 4373& 4447& 4767& 4874\\
5035& 5129& 5218& 5357& 5403& 5658& 5738& 5785& 5874& 5892\\
6061& 6063& 6578& 7024& 7030& 7429& 7627& 7719& 7793& 7940\\
7965& 7989& 8046& 8059& 8086& 8156& 8243& 8419& 8645& 8829\\
8995& 9001& 9134& 9235& 9248& 9647& 9729& 9937& 9971& 9982\\
\hline
\end{tabular}
\end{table}
\egroup

\begin{figure}
\includegraphics[width=.49\columnwidth]{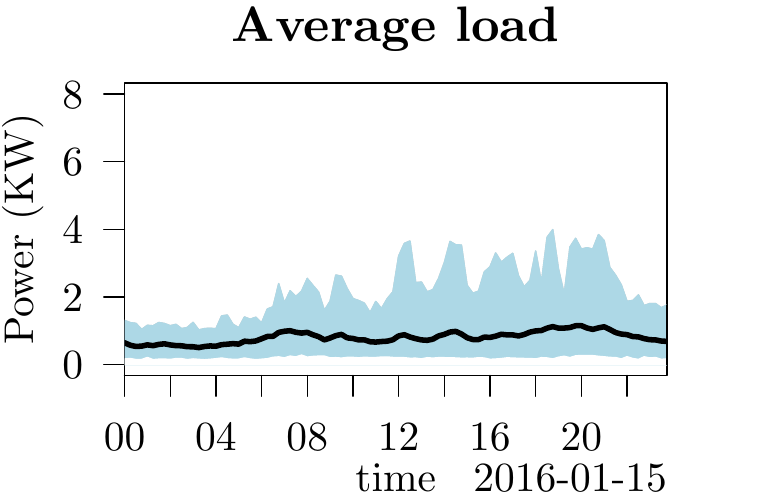}
\includegraphics[width=.49\columnwidth]{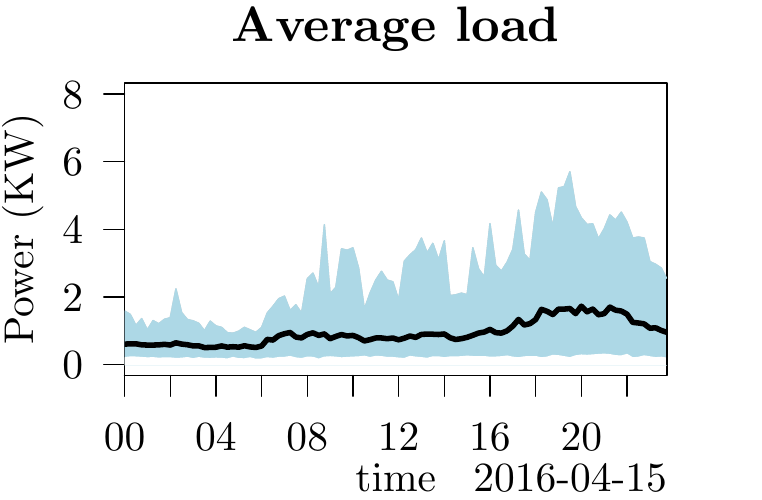}
\includegraphics[width=.49\columnwidth]{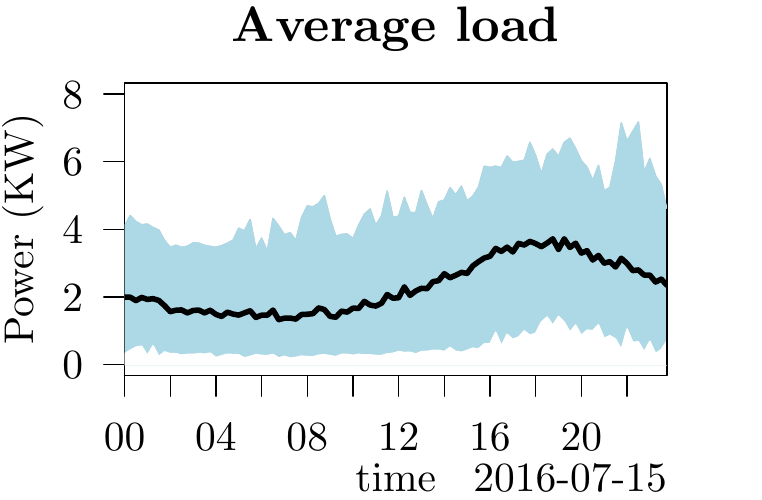}
\includegraphics[width=.49\columnwidth]{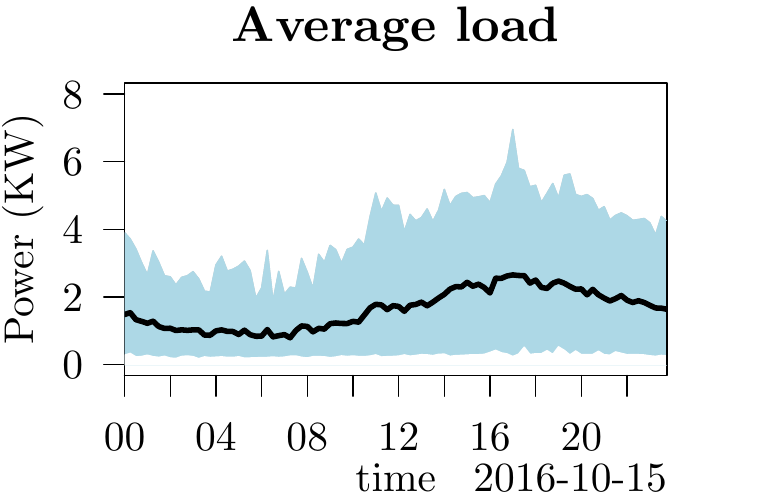}
\caption{Daily average load}
\label{fig:dailyuse}
\end{figure}

\begin{figure}
\includegraphics[width=.49\columnwidth]{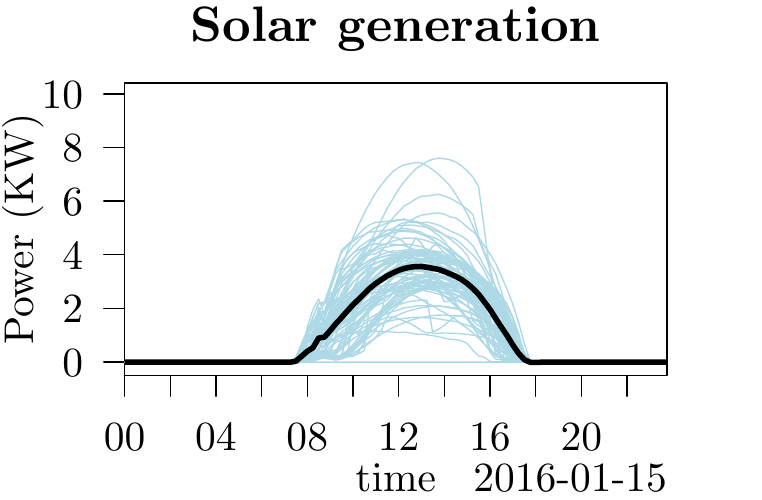}
\includegraphics[width=.49\columnwidth]{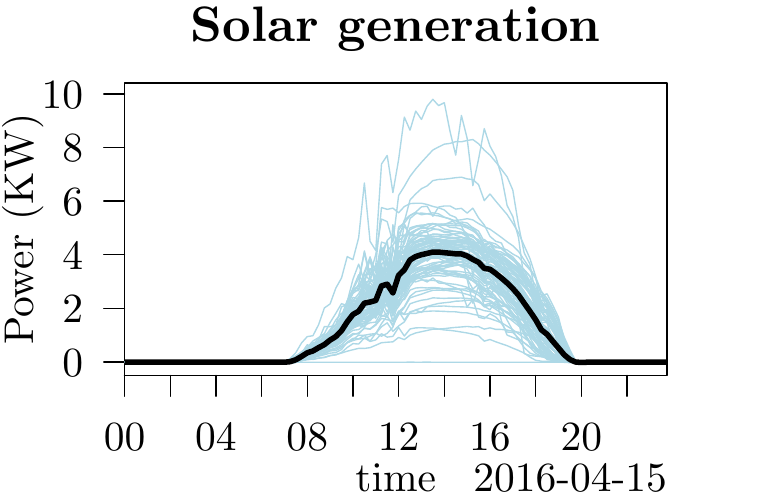}
\includegraphics[width=.49\columnwidth]{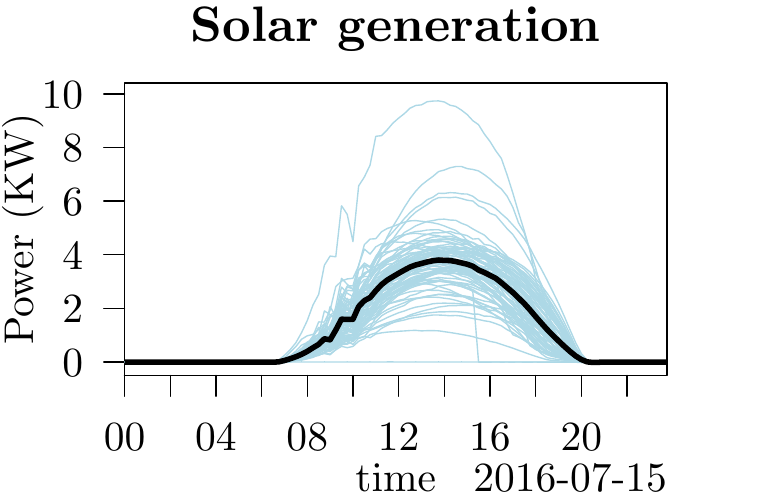}
\includegraphics[width=.49\columnwidth]{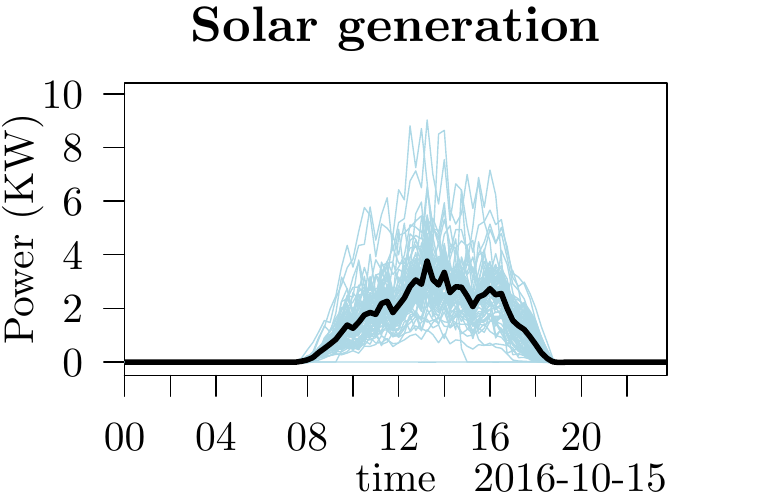}
\caption{Daily Solar PV generation}
\label{fig:dailygen}
\end{figure}

In Figure~\ref{fig:dailyuse}, we show the daily average consumption per
household in black solid line.
The shaded light blue area represents the interval
between the \num{5}\% and \num{95}\% quantiles of the consumption distribution
of the community prosumers. We have chosen four specific dates in
different seasons. These dates are 2016-01-01, 2016-04-01,
2016-07-01 and 2016-10-01. Since each prosumer has PV rooftop panel,
the daily average solar power generation per prosumer is depicted
in Figure~\ref{fig:dailygen} in black solid line for the same specific dates.
In addition, the light blue solid lines are the power generation curves
for every prosumer in the community.

The total electricity consumption of the community during 2016 is
\num{971681}~\si{\MW\hour} and the total generation is
\num{598349}~\si{\MW\hour}.
The monthly
total consumption and generation for the 80 prosumers is given in
Table~\ref{tab:totenergy}. Notice that
consumption is larger in summer and fall months because of the
use of air conditioners. In these months the solar generation
also increases but not at the same rate as that of the consumptions.
The consumption increase from January to July is \num{236}\% while
the generation increase is only \num{150}\%. The net consumption of the
community in February and March is negative.

\begin{table}
\caption{Community monthly total consumption and generation}
\label{tab:totenergy}
\begin{center}
\begin{tabular}{S[table-format=2.0]S[table-format=6.2]S[table-format=6.2]S[table-format=-6.2]}
\hline
\multicolumn{1}{c}{\textbf{Month}} &
\multicolumn{1}{c}{\textbf{(a)}} &
\multicolumn{1}{c}{\textbf{(b)}} &
\multicolumn{1}{c}{\textbf{(a)--(b)}} \\
\hline
1 &  56807.87 & 44503.73 & 12304.14 \\
2 &  48200.62 & 52105.83 & -3905.21 \\
3 &  52714.26 & 52944.47 &  -230.21 \\
4 &  60270.83 & 51398.36 &  8872.47 \\
5 &  77184.61 & 48118.61 & 29066.00 \\
6 & 113583.74 & 61418.20 & 52165.54 \\
7 & 134202.32 & 66716.79 & 67485.52 \\
8 & 119990.42 & 54610.72 & 65379.69 \\
9 & 109313.42 & 54128.38 & 55185.04 \\
10 & 83020.00 & 53773.55 & 29246.45 \\
11 & 55200.10 & 33601.74 & 21598.36 \\
12 & 61193.50 & 25028.61 & 36164.89 \\
Total & 971681.68 & 598348.98 & 373332.69 \\
\hline
\multicolumn{4}{l}{(a) Energy Consumption \si{\kW\hour}} \\
\multicolumn{4}{l}{(b) Energy Generation \si{\kW\hour}}
\end{tabular}
\end{center}
\end{table}

\begin{figure}
\includegraphics[width=\columnwidth]{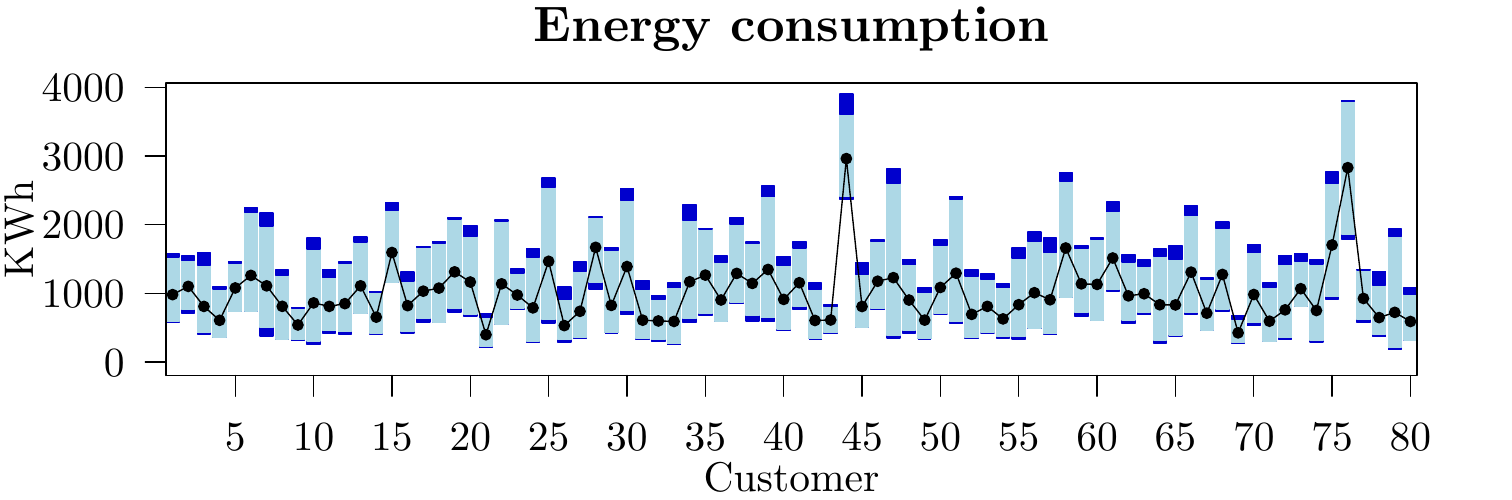}
\includegraphics[width=\columnwidth]{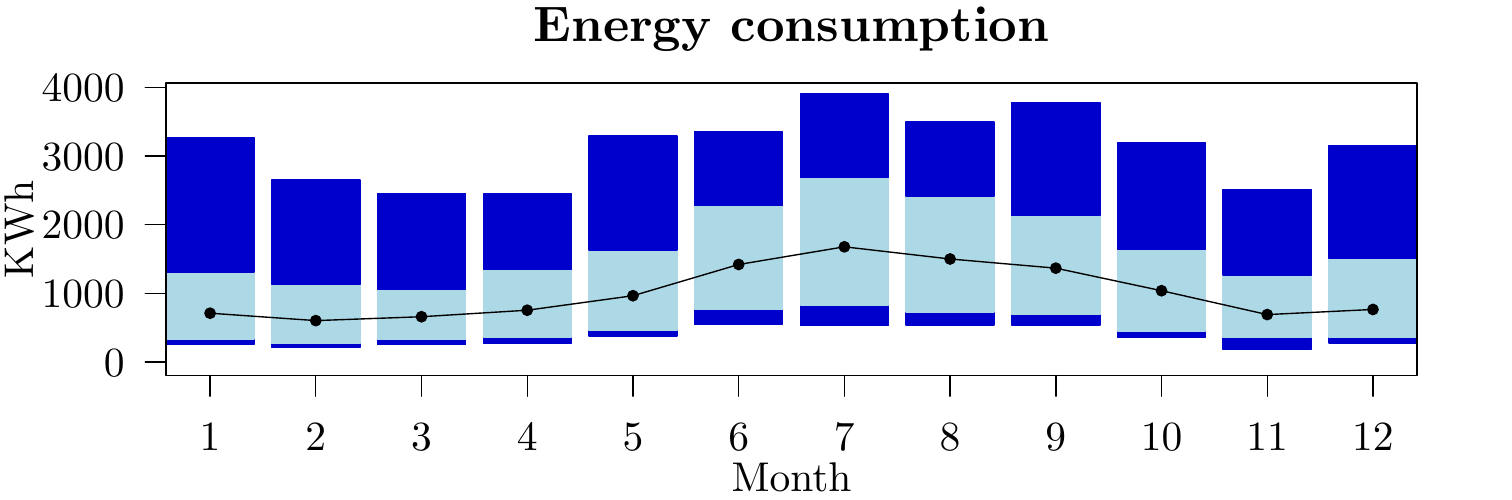}
\caption{Community load}
\label{fig:comuse}
\end{figure}

\begin{figure}
\includegraphics[width=\columnwidth]{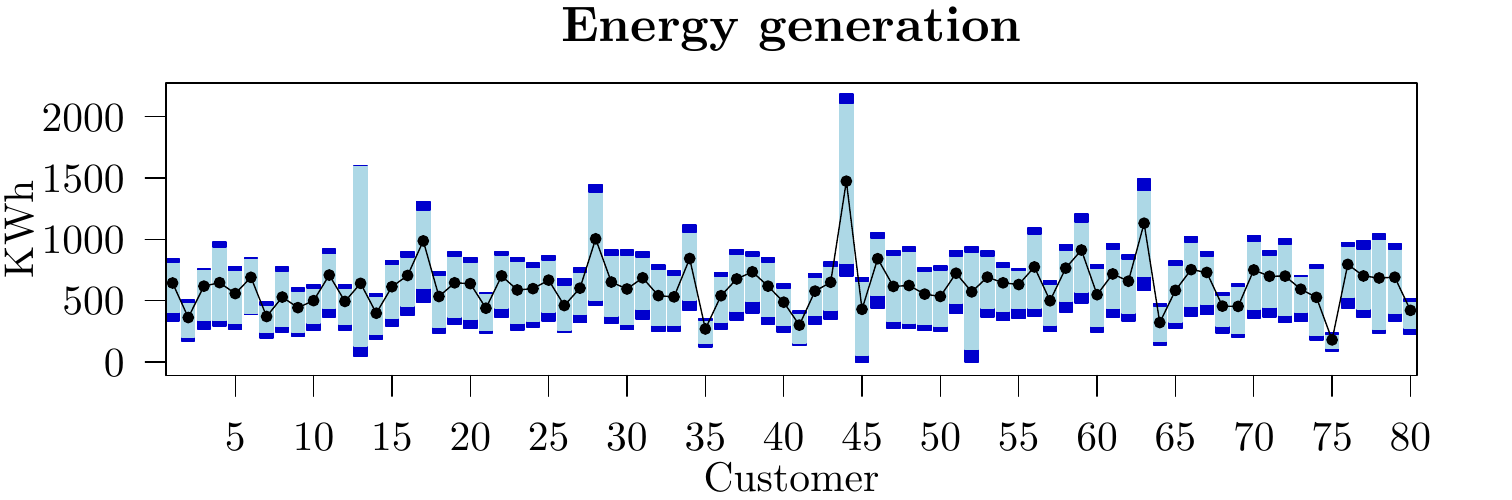}
\includegraphics[width=\columnwidth]{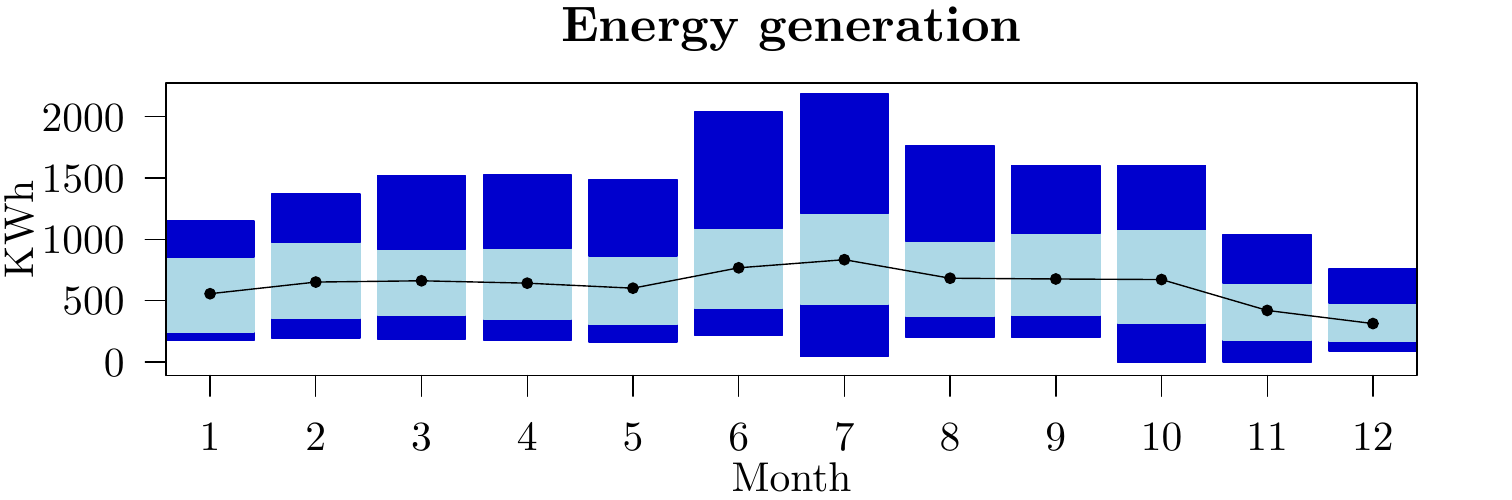}
\caption{Community PV solar generation}
\label{fig:comgen}
\end{figure}

The average monthly consumption and generation per prosumer is
\num{1012.17}~\si{\kW\hour} and \num{623.28}~\si{\kW\hour}, respectively.
The distribution of the monthly community consumption and generation
are depicted in Figures~\ref{fig:comuse} and \ref{fig:comgen}, respectively.
We show two plots for each figure. The first plot represents the
distribution of consumption and generation per prosumer, while the
second one is the distribution of consumption and generation per month.
The average value is represented by a solid black line with round marks.
The interquantile interval between \num{5}\% and \num{95}\% is shown as a light
blue bar. The remaining \num{10}\% cases are shown as dark blue bars.
We can see that most residents have monthly energy consumptions near
the average value with some seasonal variation.
There are two residents (codes 5357 and 9647 corresponding to
positions 44 and 76 in the horizontal axis)
with average monthly consumptions near to $3000~\si{\kW\hour}$, much higher
than the rest of residents.
The plots of the rooftop PV solar generation distribution show
that the monthly generation
is near to the average value of $623.28~\si{\kW\hour}$ with some seasonal variation.
As it is expected, the generation is higher during June and July because
there are more insolation hours and lower in November and December.

\subsection{Results}

We analyze now the cost of energy for each resident depending on the
billing mechanism\footnote{Notice that we are
analyzing only the cost of the consumed energy. In some European
countries the cost of electricity has a term depending
on the contracted power and possible penalties for exceeding the
contracted power. We are not considering these terms but this does not
detract from our study.}.
In both the cases, we show a table
with the sum of the costs for the \num{80} households for each month and a
figure with the distribution of the monthly cost of electricity. We
include two plots in this figure.
The first plot represents the monthly cost distribution per prosumer,
while the second one is the prosumer cost per month.
The average value is represented by a solid black line with round marks.
The interquantile interval between \num{5}\% and \num{95}\% is shown as a light
blue bar. The remaining $10\%$ cases are shown as dark blue bars.

In Table~\ref{tab:nmetsav1} we show the costs for the net
metering billing mechanism. In this case, the total
annual cost for the community is \$\num{42973.74}, corresponding to a
monthly average per household of \$\num{44.76}. If the residents decide to share
their solar rooftop generation and allocate the costs according to
Allocation~\ref{alloc:nmet} (\ref{eq:alloc1}), then the total annual
cost is \$\num{41337.22} corresponding to a monthly average cost per household
of \$\num{43.10} and a cost reduction of \num{3.96}\%.

\begin{figure}
\includegraphics[width=\columnwidth]{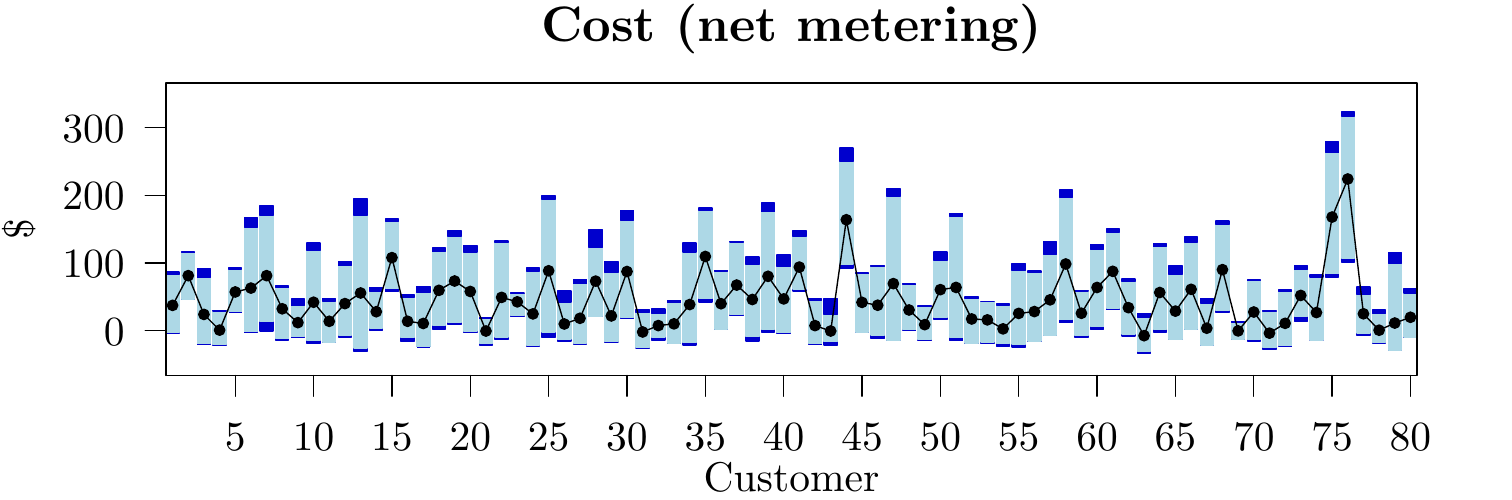}
\includegraphics[width=\columnwidth]{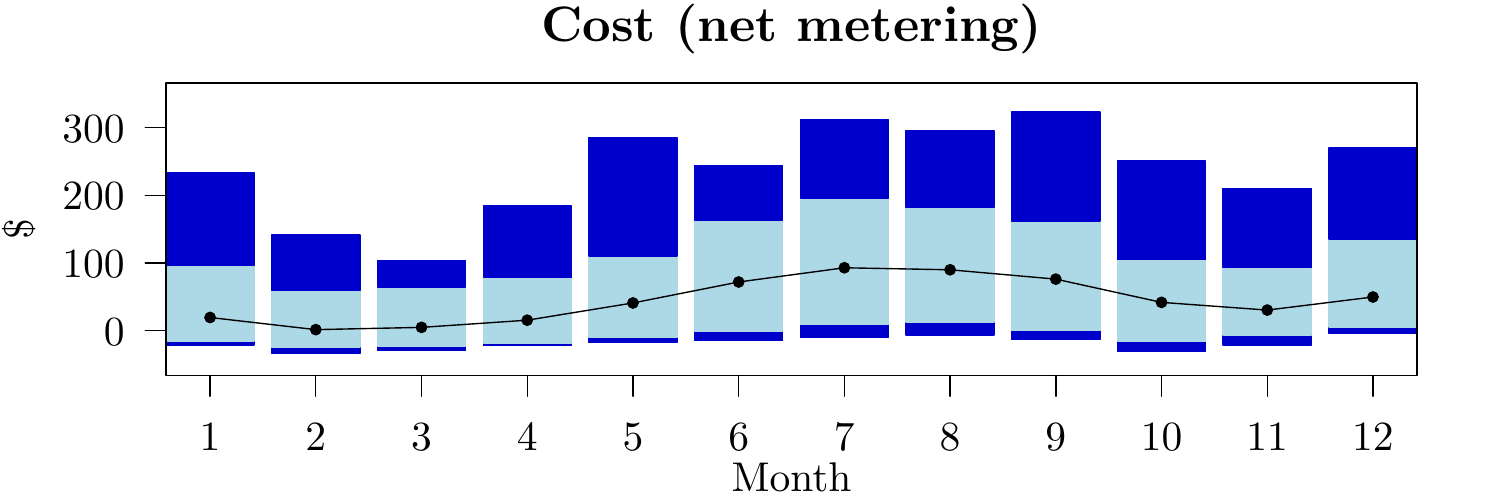}
\caption{Community cost for net metering billing mechanism}
\label{fig:nmetcos}
\end{figure}

\begin{figure}
\includegraphics[width=\columnwidth]{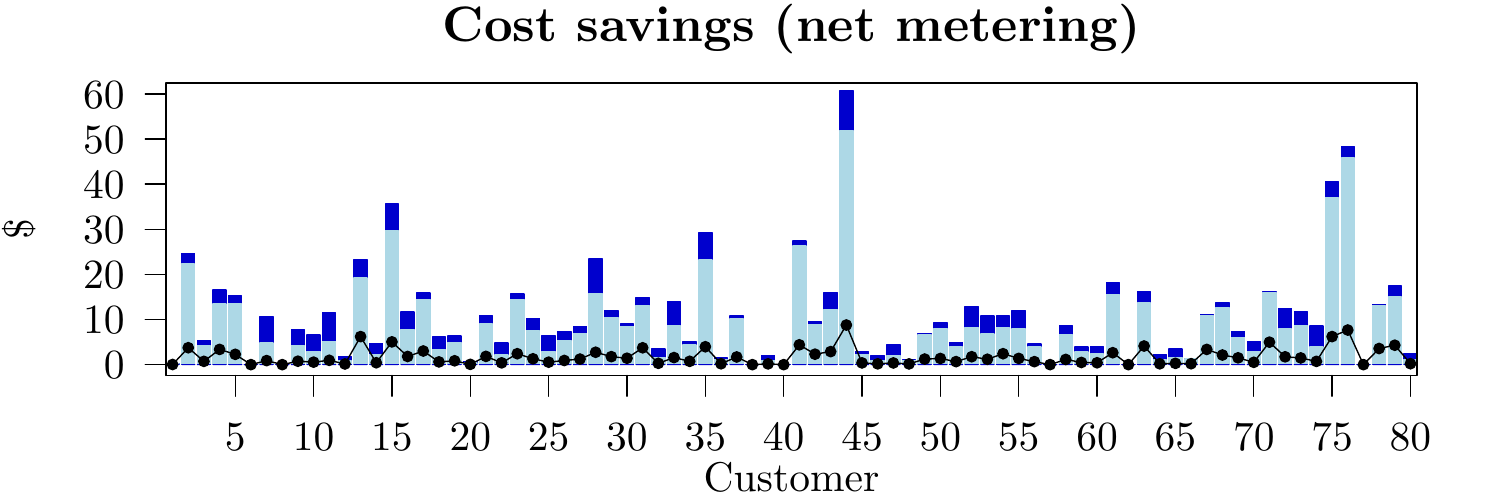}
\includegraphics[width=\columnwidth]{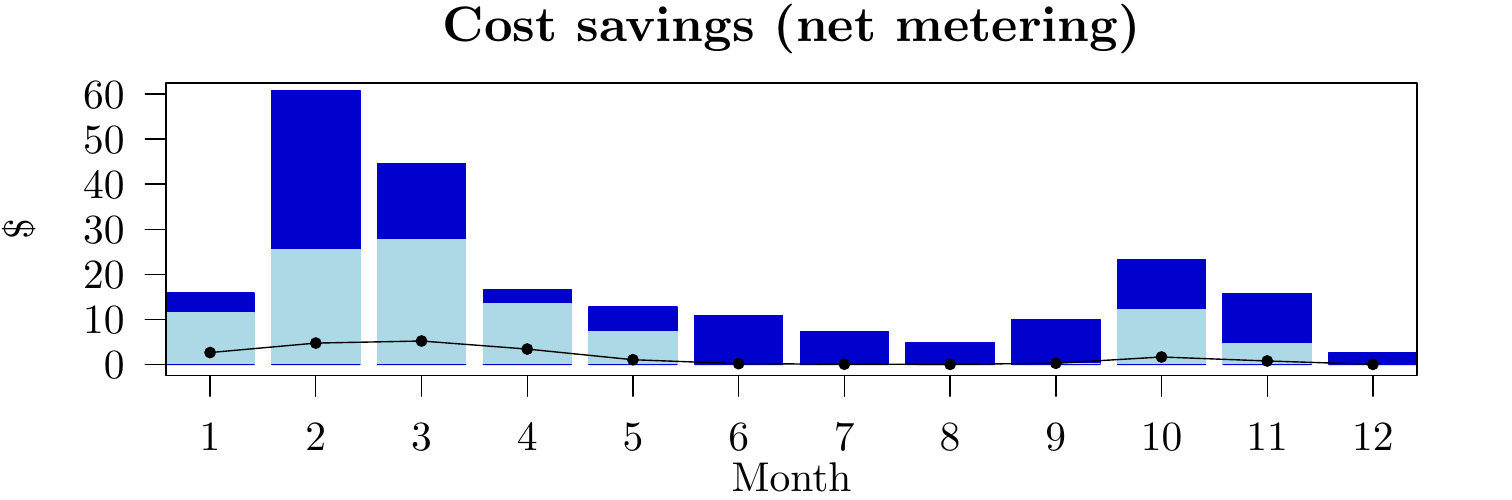}
\caption{Community savings for net metering billing mechanism}
\label{fig:nmetsav}
\end{figure}

The distribution of the monthly costs and savings are depicted in
Figures~\ref{fig:nmetcos} and \ref{fig:nmetsav}.
In Figure~\ref{fig:nmetcos} we show the monthly cost distribution,
while in Figure~\ref{fig:nmetcos} we show the monthly cost savings
distribution.
Note that the higher cost savings are obtained in
February and March. The reason is that the net consumption in these
months is negative, see Table~\ref{tab:totenergy}.
Finally, in Table~\ref{tab:nmetsav2}, we show the savings for the households.
Due to space limitation we only show the \num{20} prosumers that obtain higher
annual savings. They are ordered in decreasing order of the relative cost
differences. There are \num{19} households that obtain a reduction higher than
\num{10}\%, \num{30} households that obtain a reduction higher than \num{5}\%
and only \num{7} households that do not obtain any reduction.

\begin{table}
\caption{Summary of cost savings for net metering (I)}
\label{tab:nmetsav1}
\begin{center}
\begin{tabular}{S[table-format=2.0]S[table-format=5.2]S[table-format=5.2]S[table-format=5.2]S[table-format=5.2]}
\hline
\multicolumn{1}{c}{\textbf{Month}} &
\multicolumn{1}{c}{\textbf{(a)}} &
\multicolumn{1}{c}{\textbf{(b)}} &
\multicolumn{1}{c}{\textbf{(c)}} &
\multicolumn{1}{c}{\textbf{(d)}} \\
\hline
 1 & 1570.02 & 1355.92 & 214.11 &  13.64 \\
 2 &  136.45 & -245.30 & 381.75 & 279.78 \\
 3 &  404.29 &  -14.46 & 418.75 & 103.58 \\
 4 & 1253.37 &  977.75 & 275.62 &  21.99 \\
 5 & 3289.97 & 3203.07 &  86.90 &   2.64 \\
 6 & 5765.62 & 5748.64 &  16.97 &   0.29 \\
 7 & 7444.56 & 7436.90 &   7.66 &   0.10 \\
 8 & 7209.74 & 7204.84 &   4.89 &   0.07 \\
 9 & 6106.10 & 6081.39 &  24.71 &   0.40 \\
10 & 3358.26 & 3222.96 & 135.30 &   4.03 \\
11 & 2445.12 & 2380.14 &  64.98 &   2.66 \\
12 & 3990.26 & 3985.37 &   4.89 &   0.12 \\
Total & 42973.74 & 41337.22 & 1636.52 & 3.96 \\
\hline
\multicolumn{5}{l}{\begin{tabular}{ll}
(a) Cost without sharing (\$)  & (b) Cost with sharing (\$) \\
(c) Cost savings (a)--(b) (\$) & (d) Cost savings (c)/$|$(a)$|$ (\%)
\end{tabular}}
\end{tabular}
\end{center}
\end{table}

\begin{table}
\caption{Summary of cost savings for net metering (II)}
\label{tab:nmetsav2}
\begin{center}
\begin{tabular}{S[table-format=4.0]S[table-format=-3.2]S[table-format=-3.2]S[table-format=-3.2]S[table-format=-3.2]}
\hline
\multicolumn{1}{c}{\textbf{Prosumer}} &
\multicolumn{1}{c}{\textbf{(a)}} &
\multicolumn{1}{c}{\textbf{(b)}} &
\multicolumn{1}{c}{\textbf{(c)}} &
\multicolumn{1}{c}{\textbf{(d)}} \\
\hline
5218 &   -3.84 &  -38.55 &  34.71 & 903.19 \\
8645 &   -2.46 &  -20.54 &  18.07 & 733.38 \\
2199 &   -4.48 &  -26.48 &  22.00 & 490.63 \\
9937 &   10.15 &  -32.74 &  42.90 & 422.42 \\
 171 &   11.90 &  -28.52 &  40.42 & 339.60 \\
3456 &  -20.98 &  -65.78 &  44.81 & 213.59 \\
8995 &  -41.87 & -101.66 &  59.79 & 142.80 \\
8243 &   45.88 &    5.43 &  40.45 &  88.16 \\
7024 &   34.27 &    5.24 &  29.02 &  84.70 \\
8046 &  -86.57 & -136.06 &  49.50 &  57.18 \\
9971 &  137.10 &   85.52 &  51.59 &  37.63 \\
5129 &   91.92 &   64.14 &  27.78 &  30.22 \\
1800 &  129.89 &   93.65 &  36.24 &  27.90 \\
9001 &  132.07 &  111.33 &  20.73 &  15.70 \\
3538 &  122.99 &  104.27 &  18.73 &  15.23 \\
5874 &  108.81 &   93.97 &  14.84 &  13.63 \\
1792 &  168.54 &  146.88 &  21.66 &  12.85 \\
1283 &  670.01 &  595.46 &  74.55 &  11.13 \\
6063 &  209.11 &  188.03 &  21.08 &  10.08 \\
2945 &  120.44 &  109.41 &  11.03 &   9.16 \\
\hline
\multicolumn{5}{l}{\begin{tabular}{ll}
(a) Cost without sharing (\$), & (b) Cost with sharing (\$) \\
(c) Cost savings (a)--(b) (\$), & (d) Cost savings (c)/$|$(a)$|$ (\%)
\end{tabular}}
\end{tabular}
\end{center}
\end{table}

We conducted a similar analysis for the net purchase and sale
billing mechanism. In this case the cost savings by month are shown
in Table~\ref{tab:nmetsav1}. The annual cost for the community if each
household pay by her own net consumption is \$\num{55609.93}, corresponding to a
monthly average per household of \$\num{57.93}. If the residents decide
to share their solar rooftop generation and allocate the costs
according to Allocation~\ref{alloc:npur}
(\ref{eq:alloc2a})--(\ref{eq:alloc2b}), then the total annual
cost is \$\num{42973.74} corresponding to a monthly average cost per household
of \$\num{44.76} and a relative cost reduction of \num{22.72}\%.

\begin{figure}
\includegraphics[width=\columnwidth]{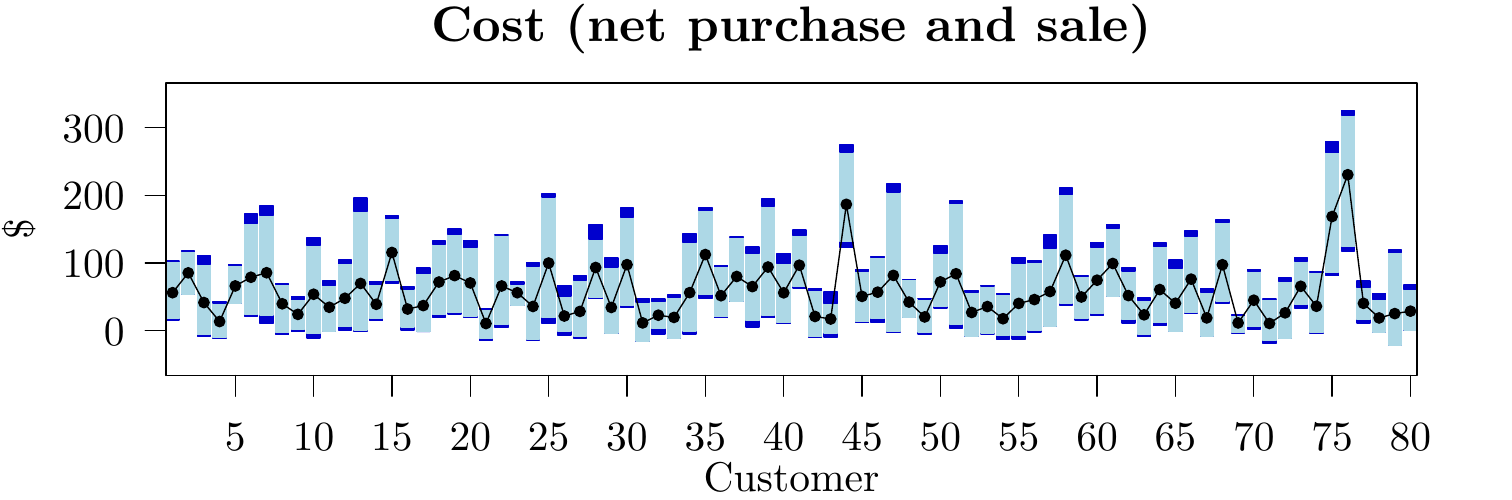}
\includegraphics[width=\columnwidth]{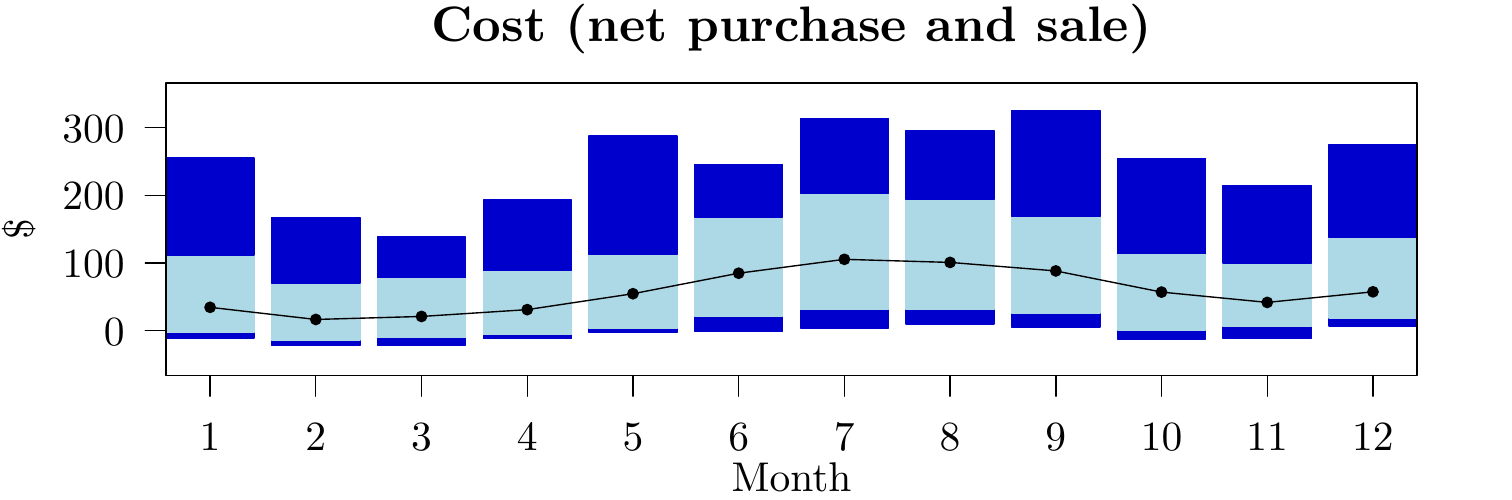}
\caption{Community cost for net purchase and sale billing mechanism}
\label{fig:npurcos}
\end{figure}

\begin{figure}
\includegraphics[width=\columnwidth]{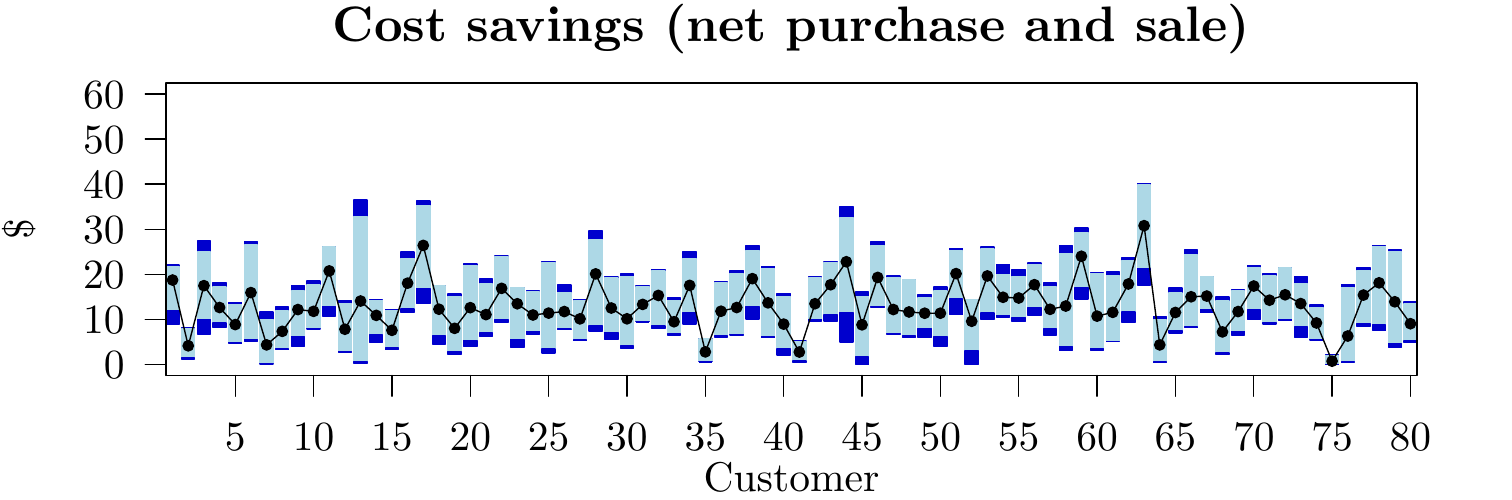}
\includegraphics[width=\columnwidth]{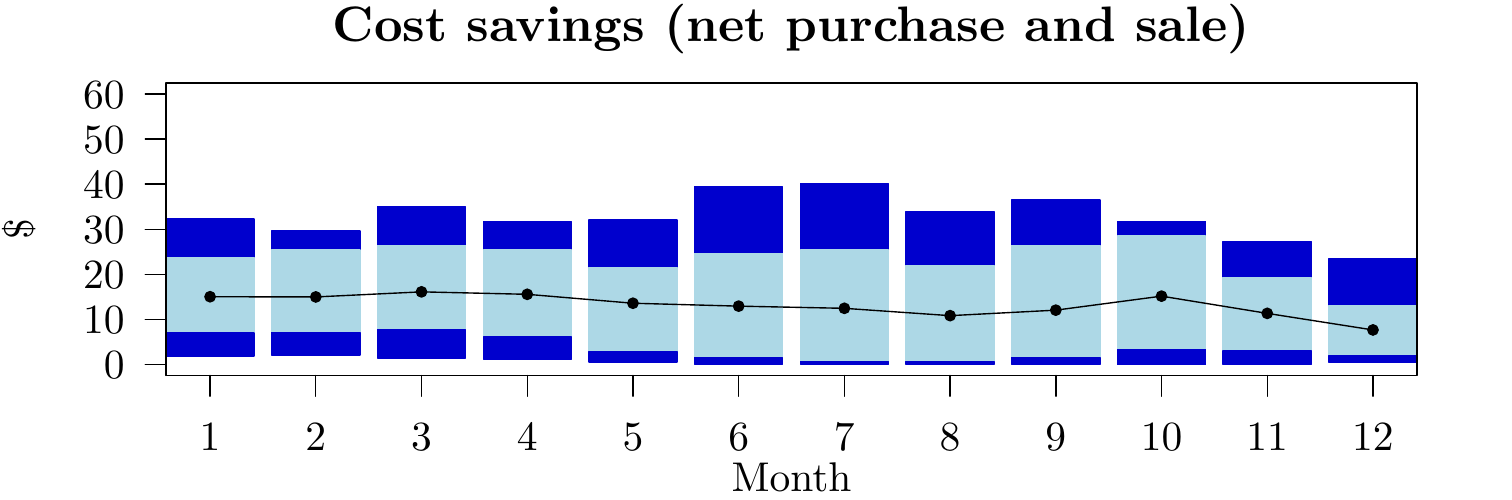}
\caption{Community savings for net purchase and sale billing mechanism}
\label{fig:npursav}
\end{figure}

The distribution of the monthly costs and savings are depicted in
Figures~\ref{fig:npurcos} and \ref{fig:npursav}. As was
analytically explained in Section~\ref{subsec:comp}- in this case the cost
savings are much higher than for the net metering billing mechanism.
Moreover, unlike that case, the higher savings are
not concentrated in two months and for a small number of prosumers.
In Table~\ref{tab:npursav2}, we show the savings for the households.
Similarly to the net metering case, we only show the
20 prosumers that obtain higher annual savings and they are ordered
in decreasing order of the relative cost savings. Every household
obtains a significant reduction of her energy cost.
There are 21 households that obtain a reduction higher than \num{50}\%,
\num{52} households obtain a reduction higher than \num{20}\%,
\num{70} higher than \num{10}\% and only one has a reduction lower than
\num{1}\%.

\begin{table}
\caption{Summary of cost savings for net purchase and sale}
\label{tab:npursav1}
\begin{center}
\begin{tabular}{S[table-format=2.0]S[table-format=5.2]S[table-format=5.2]S[table-format=-5.2]S[table-format=-2.2]}
\hline
\multicolumn{1}{c}{\textbf{Month}} &
\multicolumn{1}{c}{\textbf{(a)}} &
\multicolumn{1}{c}{\textbf{(b)}} &
\multicolumn{1}{c}{\textbf{(c)}} &
\multicolumn{1}{c}{\textbf{(d)}} \\
\hline
1 & 2773.90 & 1570.02 & 1203.88 & 43.40 \\
2 & 1336.72 & 136.45 & 1200.27 & 89.79 \\
3 & 1694.22 & 404.29 & 1289.93 & 76.14 \\
4 & 2500.15 & 1253.37 & 1246.78 & 49.87 \\
5 & 4378.49 & 3289.97 & 1088.52 & 24.86 \\
6 & 6802.61 & 5765.62 & 1036.99 & 15.24 \\
7 & 8442.62 & 7444.56 & 998.06 & 11.82 \\
8 & 8077.89 & 7209.74 & 868.15 & 10.75 \\
9 & 7071.99 & 6106.10 & 965.89 & 13.66 \\
10 & 4572.32 & 3358.26 & 1214.06 & 26.55 \\
11 & 3353.83 & 2445.12 & 908.72 & 27.09 \\
12 & 4605.20 & 3990.26 & 614.95 & 13.35 \\
Total  & 55609.93 & 42973.74 & 12636.18 & 22.72 \\
\hline
\multicolumn{5}{l}{\begin{tabular}{ll}
(a) Cost without sharing (\$)  & (b) Cost with sharing (\$) \\
(c) Cost savings (a)--(b) (\$) & (d) Cost savings (c)/$|$(a)$|$ (\%)
\end{tabular}}
\end{tabular}
\end{center}
\end{table}

\begin{table}
\caption{Summary of cost savings for net purchase and sale (II)}
\label{tab:npursav2}
\begin{center}
\begin{tabular}{S[table-format=4.0]S[table-format=3.2]S[table-format=-3.2]S[table-format=-3.2]S[table-format=-3.2]}
\hline
\multicolumn{1}{c}{\textbf{Prosumer}} &
\multicolumn{1}{c}{\textbf{(a)}} &
\multicolumn{1}{c}{\textbf{(b)}} &
\multicolumn{1}{c}{\textbf{(c)}} &
\multicolumn{1}{c}{\textbf{(d)}} \\
\hline
8995 &  129.49 &  -41.87 & 171.37 & 132.34 \\
8046 &  282.91 &  -86.57 & 369.47 & 130.60 \\
3456 &  139.16 &  -20.98 & 160.14 & 115.07 \\
2199 &  128.52 &   -4.48 & 133.00 & 103.49 \\
5218 &  208.66 &   -3.84 & 212.50 & 101.84 \\
8645 &  138.64 &   -2.46 & 141.11 & 101.78 \\
9937 &  227.65 &   10.15 & 217.50 &  95.54 \\
 171 &  163.75 &   11.90 & 151.85 &  92.73 \\
7024 &  213.38 &   34.27 & 179.12 &  83.94 \\
8243 &  228.18 &   45.88 & 182.30 &  79.89 \\
1800 &  446.96 &  129.89 & 317.07 &  70.94 \\
3482 &  276.09 &   92.10 & 183.99 &  66.64 \\
5129 &  253.90 &   91.92 & 161.98 &  63.80 \\
 781 &  416.68 &  167.39 & 249.29 &  59.83 \\
9001 &  317.79 &  132.07 & 185.72 &  58.44 \\
1792 &  385.18 &  168.54 & 216.64 &  56.24 \\
5874 &  245.45 &  108.81 & 136.64 &  55.67 \\
9971 &  304.19 &  137.10 & 167.08 &  54.93 \\
6578 &  429.52 &  193.67 & 235.85 &  54.91 \\
2945 &  261.49 &  120.44 & 141.05 &  53.94 \\
\hline
\multicolumn{5}{l}{\begin{tabular}{ll}
(a) Cost without sharing (\$)  & (b) Cost with sharing (\$) \\
(c) Cost savings (a)--(b) (\$) & (d) Cost savings (c)/$|$(a)$|$ (\%)
\end{tabular}}
\end{tabular}
\end{center}
\end{table}

We could also validate our comparative analysis between the mechanisms. Net metering produces lower costs for the households than net purchase and sale. But from the point of view
of saving cost due to sharing, net purchase and sale is the most interesting as
it promotes association by an effective sharing of the energy excesses
and producing significant reduction of the energy costs for every household.

\section{Conclusions}
\label{sec:Con}

Drastic cost reduction in the PV systems technology in the last few years has
resulted in significant increase in their worldwide installation. Attractive
billing methods implemented by different system operators have also encouraged
more houses to install PV systems. In this paper, we have explored the idea of
sharing the electricity generation by PV systems of different households among
each other and improving their profits promoting the use of clean energy more
in the power system. We considered sharing under three different
programs: feed-in tariff, net metering, and net purchase and sale. In feed-in tariff, there is no advantage in sharing. In net metering,
and net purchase and sale, sharing is advantageous if and only if the retail
price of electricity by the utility is more than the price of selling
electricity to the utility. Under that favorable sharing condition, we found
rules for allocating their joint cost based on cost causation principle. The
allocations also follow standalone cost principle - \emph{i.e.}, they are in the
core of the cooperative games of net electricity consumption cost. We have also
performed a comparative analysis between net metering, and net purchase and
sale. These results will definitely attract more households to share their PV
systems generated electricity with each other and as a result, the whole
society will benefit. We have verified our developed results in the data
set of a community of residential households in Austin, Texas. The results show
significant advantage in sharing PV systems among prosumers.
\bibliographystyle{IEEEtran}
\bibliography{Rooftop}

\appendix

\section{Proofs}
\subsection{Proof of Theorem~\ref{th:nmrsc}}

Let us consider two disjoint coalitions $\mathcal S$ and $\mathcal T$, then
\begin{align*}
C_{\mathcal S \cup \mathcal T}^{\text{NM}} = \lambda
Q_{\mathcal S \cup \mathcal T}^{\text{NM}} - \mu
G_{\mathcal S \cup \mathcal T}^{\text{NM}},
\end{align*}
where
\begin{align*}
Q_{\mathcal S \cup \mathcal T}^{\text{NM}} &=
(Q_{\mathcal S}^{\text{NM}} + Q_{\mathcal T}^{\text{NM}} -
G_{\mathcal S}^{\text{NM}} - G_{\mathcal T}^{\text{NM}})_+, \\
G_{\mathcal S \cup \mathcal T}^{\text{NM}} &=
(G_{\mathcal S}^{\text{NM}} + G_{\mathcal T}^{\text{NM}} -
Q_{\mathcal S}^{\text{NM}} - Q_{\mathcal T}^{\text{NM}})_+,
\end{align*}
and
\begin{align*}
C_{\mathcal S}^{\text{NM}} + C_{\mathcal T}^{\text{NM}} &=
\lambda (Q_{\mathcal S}^{\text{NM}} + Q_{\mathcal T}^{\text{NM}}) -
\mu (G_{\mathcal S}^{\text{NM}} + G_{\mathcal T}^{\text{NM}}).
\end{align*}

In net metering mechanism $Q_{\mathcal S}^{\text{NM}}\geq 0$, $G_{\mathcal S}^{\text{NM}}\geq 0$,
and one of them is always zero,
\emph{i.e.} $Q_{\mathcal S}^{\text{NM}}G_{\mathcal S}^{\text{NM}}=0$ for
any coalition $\mathcal S \subseteq \mathcal N$.
Then we can distinguish four possible cases:

Case (\textsc{i}): $G_{\mathcal S}^{\text{NM}}=G_{\mathcal T}^{\text{NM}}=0$,
\begin{align*}
C_{\mathcal S \cup \mathcal T}^{\text{NM}} = \lambda
(Q_{\mathcal S}^{\text{NM}} + Q_{\mathcal T}^{\text{NM}}) =
C_{\mathcal S}^{\text{NM}} + C_{\mathcal T}^{\text{NM}}.
\end{align*}

Case (\textsc{ii}): $Q_{\mathcal S}^{\text{NM}}=Q_{\mathcal T}^{\text{NM}}=0$,
\begin{align*}
C_{\mathcal S \cup \mathcal T}^{\text{NM}} = -\mu
(G_{\mathcal S}^{\text{NM}} + G_{\mathcal T}^{\text{NM}}) =
C_{\mathcal S}^{\text{NM}} + C_{\mathcal T}^{\text{NM}}.
\end{align*}

Case (\textsc{iii}): $G_{\mathcal S}^{\text{NM}}=Q_{\mathcal T}^{\text{NM}}=0$,
\begin{align*}
C_{\mathcal S \cup \mathcal T}^{\text{NM}} &= \lambda
(Q_{\mathcal S}^{\text{NM}} - G_{\mathcal T}^{\text{NM}})_+ - \mu
(G_{\mathcal T}^{\text{NM}} - Q_{\mathcal S}^{\text{NM}})_+, \\
C_{\mathcal S}^{\text{NM}} + C_{\mathcal T}^{\text{NM}} &= \lambda
Q_{\mathcal S}^{\text{NM}} - \mu G_{\mathcal T}^{\text{NM}}.
\end{align*}

We have two cases, either $Q_{\mathcal S}^{\text{NM}} \geq G_{\mathcal T}^{\text{NM}}$ or
$Q_{\mathcal S}^{\text{NM}} < G_{\mathcal T}^{\text{NM}}$. In both cases,
$C_{\mathcal S \cup \mathcal T}^{\text{NM}} \leq
C_{\mathcal S}^{\text{NM}} + C_{\mathcal T}^{\text{NM}}$ if and only if $\mu \leq \lambda$.

Case (\textsc{iv}): $Q_{\mathcal S}^{\text{NM}}=G_{\mathcal T}^{\text{NM}}=0$,
\begin{align*}
C_{\mathcal S \cup \mathcal T}^{\text{NM}} &= \lambda
(Q_{\mathcal T}^{\text{NM}} - G_{\mathcal S}^{\text{NM}})_+ - \mu
(G_{\mathcal S}^{\text{NM}} - Q_{\mathcal T}^{\text{NM}})_+, \\
C_{\mathcal S}^{\text{NM}} + C_{\mathcal T}^{\text{NM}} &=
\lambda Q_{\mathcal T}^{\text{NM}} - \mu G_{\mathcal S}^{\text{NM}}.
\end{align*}

We have two cases, either $Q_{\mathcal T}^{\text{NM}} \geq G_{\mathcal S}^{\text{NM}}$ or
$Q_{\mathcal T}^{\text{NM}} < G_{\mathcal S}^{\text{NM}}$. In both cases,
$C_{\mathcal S \cup \mathcal T}^{\text{NM}} \leq
C_{\mathcal S}^{\text{NM}} + C_{\mathcal T}^{\text{NM}}$ if and only if $\mu \leq \lambda$.

Thus, subadditivity of the cost function $C^{\text{NM}}$ of the cooperative
game $(\mathcal N,C^{\text{NM}})$ is equivalent to $\lambda \geq \mu$.

\subsection{Proof of Theorem~\ref{th:nmrcc}}

According to Definition~\ref{def:CCBA}, we have to
prove that the cost allocation rule given by (\ref{eq:alloc1}) satisfies
the Axioms~\ref{ax:Eq}--\ref{ax:BB} and \ref{ax:cc1}--\ref{ax:cc2}.
Let $D_i=Q_i^{\text{NM}}-G_i^{\text{NM}}$ be the net consumption. The allocation follows the
following axioms:

Axiom~\ref{ax:Eq} (Equity):
It is easy to see that if two households $i$ and $j$ have same net consumptions,
the allocated costs must be the same \emph{i.e.}, if
$D_i = D_j$ then
$x_i^{\text{NM}}=x_j^{\text{NM}}$.

Axiom~\ref{ax:Mo} (Monotonicity):
Under the cost allocation rule (\ref{eq:alloc1}),
if $D_iD_j\geq 0$ and $|D_i| \geq |D_j|$ then $|x_i^{\text{NM}}| \geq |x_j^{\text{NM}}|$
and this proves Monotonicity.

Axiom~\ref{ax:IR} (Individual Rationality):
\begin{align}\label{eq:icost}
C_i^{\text{NM}} &=
\left\{
\begin{array}{ll}
\lambda Q_i^{\text{NM}}, & \mathrm{if} \ D_i \geq 0,\\
 - \mu G_i^{\text{NM}}, & \mathrm{if} \ D_i < 0.
\end{array}
\right.
\end{align}

Since $\lambda \geq \mu$, comparing (\ref{eq:alloc1})
with (\ref{eq:icost}) we can say that
the allocated cost will be less than the net consumption cost
if the household would not have joined the aggregation
\emph{i.e.}, $x_i^{\text{NM}}\leq C_i^{\text{NM}}$.

Axiom~\ref{ax:BB} (Budget Balance):

If $D_{\mathcal N} \geq 0$:
\begin{align*}
\sum_{i\in\mathcal N} x_i^{\text{NM}} &= + \sum_{i\in\mathcal N} \lambda (Q_i^{\text{NM}}-G_i^{\text{NM}}) = C_{\mathcal N}^{\text{NM}},
\end{align*}

If $D_{\mathcal N}<0$:
\begin{align*}
\sum_{i\in\mathcal N} x_i^{\text{NM}} &= -
\sum_{i\in\mathcal N} \mu (G_i^{\text{NM}}-Q_i^{\text{NM}}) = C_{\mathcal N}^{\text{NM}}.
\end{align*}

So the cost allocation rule is such that the sum of allocated costs are equal
to the total net electricity consumption cost \emph{i.e.},
\begin{align*}
\sum_{i \in \mathcal N} x_i^{\text{NM}}= C_{\mathcal N}^{\text{NM}}.
\end{align*}

Recall from Definition~\ref{def:CCM} that a household $i$ with positive net
consumption $D_i=Q_i^{\text{NM}}$ is causing cost to the system and with
negative $D_i=-G_i^{\text{NM}}$ is mitigating cost to the system.

Axiom~\ref{ax:cc1} (Penalty for causing cost):
From (\ref{eq:alloc1}), if $D_i = Q_i^{\text{NM}}\geq 0 $, $x_i^{\text{NM}} \geq 0$
\emph{i.e.}, those individuals causing cost will pay for it.

Axiom~\ref{ax:cc2} (Reward for cost mitigation):
From (\ref{eq:alloc1}), if $D_i = -G_i^{\text{NM}} < 0 $, $x_i^{\text{NM}} < 0 $
\emph{i.e.}, those individuals mitigating cost will be rewarded. The rate of
penalty and reward is same here.

And we conclude that the cost allocation defined by (\ref{eq:alloc1})
is a cost causation based allocation.

\subsection{Proof of Theorem~\ref{th:nmrc}}

In order to prove that the cost allocation rule
satisfies the standalone principle (Axiom~\ref{ax:ST}),
two cases are considered:

If $D_{\mathcal N} \geq 0$:
\begin{align*}
\sum_{i \in S} x_i^{\text{NM}} &= \lambda D_{\mathcal S}, \\
C_{\mathcal S}^{\text{NM}} &=+\lambda Q_{\mathcal S}^{\text{NM}}, \quad
\mathrm{if} \ D_{\mathcal S} \geq 0, \\
C_{\mathcal S}^{\text{NM}} &=-\mu G_{\mathcal S}^{\text{NM}},
\quad \mathrm{if} \ D_{\mathcal S} \leq 0.
\end{align*}

If $D_{\mathcal N} \leq 0$:
\begin{align*}
\sum_{i \in S} x_i^{\text{NM}} &= \mu D_{\mathcal S}, \\
C_{\mathcal S}^{\text{NM}} &=+\lambda Q_{\mathcal S}^{\text{NM}}, \quad
\mathrm{if} \ D_{\mathcal S} \geq 0, \\
C_{\mathcal S}^{\text{NM}} &=-\mu G_{\mathcal S}^{\text{NM}},
\quad \mathrm{if} \ D_{\mathcal S} \leq 0.
\end{align*}

So from above, we can conclude that for every aggregation
$\mathcal S \subseteq \mathcal{N}$:
\begin{align*}
\sum_{i \in S} x_i^{\text{NM}}\leq
C_{\mathcal S}^{\text{NM}},
\end{align*}
and this proves that the cost allocation satisfies
the standalone cost principle.

\subsection{Proof of Theorem~\ref{th:npssc}}
The cooperative game
$(\mathcal N,C^{\text{NPS}})$ for the billing period $[t_0,t_f]$
is subadditive if and only if the game
$(\mathcal N,C^{\text{NPS}}(t))$ for $t \in [t_0,t_f]$ is subadditive.
Since the instantaneous game $(\mathcal N,C^{\text{NPS}}(t))$ is equivalent
to a net metering coalitional game for a billing period
reduced to the time instant $t$, then by Theorem~\ref{th:nmrsc},
subadditivity is equivalent to $\lambda \geq \mu$.

\subsection{Proof of Theorem~\ref{th:npscc}}
The cost allocation $x_i^{\text{NPS}}(t)$ given by
(\ref{eq:alloc2b}) satisfies Axioms~{1--7} for the instantaneous cooperative
game $(\mathcal N,C^{\text{NPS}}(t))$ where $t \in [t_0,t_f]$ because
the game $(\mathcal N,C^{\text{NPS}}(t))$ is equivalent to
a net metering cooperative game where the billing interval reduces to
the time instant $t$. Thus, $x_i^{\text{NPS}} = \int_{t_0}^{t_f} x_i^{\text{NPS}}(t) dt$ is a cost
causation based cost allocation that belongs to the core of the
cooperative game $(\mathcal N,C^{\text{NPS}})$ for the
net purchase and sale program.

\subsection{Proof of Theorem~\ref{th:difcost}}

We consider two cases.

Case $D_{\mathcal S} \geq 0$: There is nonnegative net consumption, then
$D_{\mathcal S}=Q_{\mathcal S}^{\text{NM}}=
Q_{\mathcal S}^{\text{NPS}}-G_{\mathcal S}^{\text{NPS}}$.
The cost
of the coalition consumption under Net metering and net purchase and
sale programs are:
\begin{align*}
C_{\mathcal S}^{\text{NM}} &=
\lambda D_{\mathcal S} =
\lambda Q_{\mathcal S}^{\text{NPS}} - \lambda G_{\mathcal S}^{\text{NPS}}, \\
C_{\mathcal S}^{\text{NPS}} &=
\lambda Q_{\mathcal S}^{\text{NPS}} - \mu G_{\mathcal S}^{\text{NPS}}.
\end{align*}
Thus,
\begin{align*}
C_{\mathcal S}^{\text{NPS}} -
C_{\mathcal S}^{\text{NM}} &=
(\lambda-\mu) G_{\mathcal S}^{\text{NPS}}.
\end{align*}

Case $D_{\mathcal S} < 0$: There is positive net generation, then
$D_{\mathcal S}=-G_{\mathcal S}^{\text{NM}}=
Q_{\mathcal S}^{\text{NPS}}-G_{\mathcal S}^{\text{NPS}}$.
The cost
of the coalition consumption under net metering and net purchase and
Sale programs are:
\begin{align*}
C_{\mathcal S}^{\text{NM}} &=
\mu D_{\mathcal S} =
\mu Q_{\mathcal S}^{\text{NPS}} - \mu G_{\mathcal S}^{\text{NPS}}, \\
C_{\mathcal S}^{\text{NPS}} &=
\lambda Q_{\mathcal S}^{\text{NPS}} - \mu G_{\mathcal S}^{\text{NPS}}.
\end{align*}
Thus,
\begin{align*}
C_{\mathcal S}^{\text{NPS}} -
C_{\mathcal S}^{\text{NM}} &=
(\lambda-\mu) Q_{\mathcal S}^{\text{NPS}}.
\end{align*}

\end{document}